\documentclass[showpacs,aps,prd,twocolumn,superscriptaddress,nofootinbib,longbibliography]{revtex4-2}

\usepackage[bookmarks, linktocpage, colorlinks = true, linkcolor = blue, urlcolor  = blue, citecolor = blue, anchorcolor = green, hyperindex = true, hyperfigures]{hyperref}

\usepackage{graphicx} 
\usepackage{amssymb}
\usepackage{dcolumn}
\usepackage{multirow}
\usepackage{amsmath,amssymb}
\usepackage{slashed}
\usepackage[usenames]{color}
\usepackage{float}

\newcommand{\calX}{\mathcal{X}}
\newcommand{\calY}{\mathcal{Y}}
\newcommand{\bk}{\boldsymbol{k}}
\newcommand{\bq}{\boldsymbol{q}}

\newcommand{\kbu}{k_{\rm bu}}
\newcommand{\ksh}{k_{\rm sh}}
\newcommand{\qbu}{q_{\rm bu}}
\newcommand{\qsh}{q_{\rm sh}}
\newcommand{\Lhat}{\hat{L}}

\newcommand{\nsat}{n_{\mathrm{sat}}}

\newcommand{\vk}{\boldsymbol{k}}

\begin{document}

\preprint{INT-PUB-24-056, RIKEN-iTHEMS-Report-24}

\title{Evolution of strangeness and hyperons in quarkyonic matter}

\author{Yuki~Fujimoto}
\email{fujimotophy@gmail.com}
\affiliation{Institute for Nuclear Theory, University of Washington, Box 351550, Seattle, WA 98195, USA}
\affiliation{Interdisciplinary Theoretical and Mathematical Sciences Program (iTHEMS), RIKEN, Wako 351-0198, Japan}
\affiliation{Department of Physics, University of California, Berkeley, CA 94720, USA}

\author{Toru~Kojo}
\email{torukojo@post.kek.jp}
\affiliation{Theory Center, IPNS, High Energy Accelerator Research Organization (KEK), 1-1 Oho, Tsukuba, Ibaraki 305-0801, Japan}
\affiliation{Department of Physics, Tohoku University, Sendai 980-8578, Japan}

\author{Larry~McLerran}
\email{mclerran@me.com}
\affiliation{Institute for Nuclear Theory, University of Washington, Box 351550, Seattle, WA 98195, USA}

\date{\today}

\begin{abstract}
We study the evolution of matter composition from nuclear to quark densities in the confining regime,
by extending an ideal model of Quarkyonic matter, IdylliQ model, to multi-flavor systems including strangeness.
The model provides a dual description of quark and baryon occupation probabilities which are determined by minimizing the energy of the system.
Saturation of low-momentum quark states drives the formation of quark matter and constrains baryon distributions, inducing statistical repulsion among baryon species.
Applying the model to charge-neutral matter composed of neutrons, $\Lambda_0$, and $\Sigma_0$ hyperons,
we find that, for typical size of baryons, $d$-quark saturation occurs before hyperons appear, 
delaying their onset and shifting the threshold density from $\sim 2$--$3n_{\rm sat}$ to $\sim 5$--$6n_{\rm sat}$ ($n_{\rm sat} \approx 0.16\,{\rm fm^{-3}}$: nuclear saturation density).
After hyperons emerge, low-momentum hyperon states remain only sparsely occupied due to the quark saturation.
These features mitigate the hyperon puzzle, in which the appearance of hyperons softens neutron star equations of state significantly 
by increasing energy density with little pressure increase. 
Our results highlight the key role of quark saturation in dense baryonic matter 
and provide new insights into the interplay between quark dynamics and hyperon physics in neutron stars.
\end{abstract}

\maketitle

\section{Introduction}

Describing the evolution of matter from hadronic to quark matter is one of the most important challenges in quantum chromodynamics (QCD) and modern nuclear physics.
The problem includes the evolution of matter compositions such as baryon species and lepton fraction.
In neutron star phenomenology, the understanding of matter content at baryon densities above $n_B \gtrsim 2\nsat$ ($\nsat \approx 0.16\,\text{fm}^{-3}$: nuclear saturation density) is crucial to delineate not only the equation of state (EoS) but also transport phenomena such as cooling. 
The need for stiffening of the EoS at a relatively low density, around $1.5~\nsat$, is established phenomenologically~\cite{Drischler:2020fvz}.
For the recent development at $n_B \lesssim 2 \nsat$, see Ref.~\cite{Drischler:2021kxf} for instance.

To bridge nuclear and quark matter descriptions, seminal works have elaborated Quarkyonic matter models, applying them to neutron-star phenomenology~\cite{McLerran:2018hbz} and also to nuclear matter~\cite{Koch:2024zag, McLerran:2024rvk}.
Quarkyonic matter~\cite{McLerran:2007qj} is a matter composed of the quark Fermi sea with the baryonic Fermi surface~\cite{McLerran:2007qj, Hidaka:2008yy, Andronic:2009gj, Kojo:2009ha, Kojo:2011cn}, and color confinement remains important until the color screening caused by quarks cuts off confining gluons.
Quarks near the Fermi surface are trapped into baryons, while the bulk of the Fermi sea is filled equally by quark states at different colors and hence is color singlet.
The range of the quark chemical potential $\mu$ for Quarkyonic matter descriptions is estimated to be $\Lambda_{\rm QCD} \lesssim \mu \lesssim \sqrt{N_c} \Lambda_{\rm QCD}$~\cite{McLerran:2007qj} with $\Lambda_{\rm QCD}$ and $N_c$ being the typical non-perturbative scale in QCD and the number of colors, respectively.

Recently, the present authors have constructed an ideal model of Quarkyonic matter, IdylliQ model~\cite{Fujimoto:2023mzy}, which allows us dual descriptions for the occupation probabilities of states for quarks and baryons.
The advantage of this dual model is that one can keep track of both degrees of freedom, which are dual to each other, all the way from nuclear to quark matter densities.
The quark matter formation is characterized by the saturation of quark states which begins at low momenta, triggering the rapid stiffening of the EoS, i.e., rapid increase in the pressure as a function of the energy density.
In the dual baryonic language, the quark Pauli blocking constraint leads to strong suppression of the baryon occupation densities at low momenta while baryons at higher momenta are free of such constraint;
the emerging picture is a momentum space shell structure of baryons
at the Fermi surface with the bulk part of the Fermi sea filled by quarks~\cite{McLerran:2018hbz, McLerran:2007qj, Jeong:2019lhv}, as originally conjectured in Ref.~\cite{McLerran:2007qj}.

The previous IdylliQ model is restricted to isospin symmetric matter with equal amount of protons and neutrons.
In this paper we extend our previous study to three-flavor matter including strange quarks, elaborating the framework to describe the evolution of matter composition.
This extension clarifies the importance of the quark physics in baryonic matter more vividly than in the flavor symmetric case.
In particular, the quark Pauli blocking alone induces statistical repulsions among different baryon species.
For simplicity and illustration of key ideas, in this paper we focus on charge neutral matter composed of charge neutral baryons, neutron $n (udd)$, the isosinglet $\Lambda_0$ $(uds, I=0)$, and a member of the isotriplet $\Sigma_0$ $(uds, I=1)$.
Below we collectively denote $\Lambda_0$ and $\Sigma_0$ as $Y(uds)$, and ignore isospin splitting in the mass, e.g., $M_Y \equiv M_{\Lambda^0} \approx M_{\Sigma^0} $.
Increasing baryon density, $d$-quark states are first saturated, imposing important constraints
on the $\Lambda_0$ and $\Sigma_0$ hyperons.
We also examine the appearance of $\Xi_0 (uss)$ which is free of the saturated $d$-quark Fermi sea.

Our descriptions should have important implications on the hyperon puzzle, if there is one, which states that the appearance of hyperons softens neutron star EoS significantly~\cite{Glendenning:1991es, Schaffner:1995th, Balberg:1997yw}, violating the observational constraints~\cite{Demorest:2010bx, Antoniadis:2013pzd, NANOGrav:2019jur}.
Hyperons, which are nonrelativistic right above their thresholds, increases the pressure scarcely but the energy density significantly.
In addition, there are a plethora of hyperon species, and the EoS becomes softer at every threshold for each hyperon.

An important question is at which density hyperons begin to appear (see, e.g., Refs.~\cite{Tolos:2020aln, Burgio:2021vgk} for reviews) and whether they appear in the cores of neutron stars.
Typical estimates based on nuclear matter models suggest its onset to be $n_B \sim 2$--$3~ \nsat$,
the density well achieved in neutron star cores.
This estimate is sensitive to nuclear repulsion that enhances the neutron chemical potential $\mu_n$. 
The onset density of hyperons is lowered for stronger nuclear repulsion since neutrons on the Fermi surface, with the energy $\mu_n  \ge M_Y$ ($M_Y$: hyperon mass), undergo weak decays into hyperons $Y$ \cite{Baldo:1999rq}.
To temper such decays, nuclear matter models must accompany repulsive $N$-$Y$ and $Y$-$Y$ interactions whose trends are not simple.
Recent lattice-QCD suggests that the $N$-$Y$ interaction is weaker than $N$-$N$ interactions~\cite{Beane:2012ey, Inoue:2018axd}, and the resultant hyperon single-particle potentials $U_Y (k)$ in nuclear matter are $U_\Lambda(0) \approx -28~\text{MeV}$ and $U_\Sigma(0) \approx 15~\text{MeV}$~\cite{Inoue:2018axd} (see also the estimate from $\Lambda$-hypernuclei data~\cite{Yamamoto:1988qz}), implying that $N$-$\Lambda$ and $N$-$\Sigma$ interaction on average are weakly attractive and repulsive, respectively.
In addition, the $YNN$ three-body interactions, which have been constrained by $\Lambda$-hypernuclei with varying atomic numbers~\cite{Yamamoto:1988qz, Guo:2021vsx, Jinno:2023xjr}, are also frequently employed \cite{Haidenbauer:2016vfq, Gerstung:2020ktv,Lonardoni:2014bwa}, although the extrapolation of models constrained at $n_B \sim n_{\rm sat}$ to the domain beyond $n_B\sim 2n_{\rm sat}$ remains to be validated.

Another direction is to assume that the strangeness enters through the emergence of strange quarks in quark matter, rather than as hyperons in baryonic matter.
In the former, the strange quark appears when $\mu$ exceeds the strange quark mass, $\mu \gtrsim M_s \approx 500$ MeV or $\mu_B \gtrsim 3M_s \approx 1500$ MeV, which is substantially larger than the masses of $\Lambda_0$ and $\Sigma_0$.
Moreover, the strange quarks in quark matter increases the pressure much faster than hyperons in baryonic matter, so that the appearance of strangeness in quark matter is not as problematic as in baryonic matter.
The quark matter inside neutron stars should be realized either by a phase transition at sufficiently low density \cite{Weissenborn:2011qu, Bonanno:2011ch, Lastowiecki:2011hh, Klahn:2013kga} or by a continuous crossover~\cite{Masuda:2012kf, Masuda:2012ed, Masuda:2015kha, Kojo:2014rca, Baym:2017whm, Baym:2019iky, Kojo:2021wax, Yamamoto:2023osc, Yamamoto:2024mta, Rijken:2024bce, Rijken:2024hmd}.
The major question is how to connect the quark matter models to baryonic matter models.

In our Quarkyonic matter descriptions for charge neutral $n$-$\Lambda_0$-$\Sigma_0$ matter, we find that the quark saturation effects introduce new qualitative perspectives into the in-medium $N$-$Y$ repulsion and mild softening associated with the appearance of the strangeness.

In this paper, we focus on the neutron-rich regime where $d$-quark states are saturated and then examine the ramification of such quark saturation near the hyperon threshold.
The main results of this paper are two-fold:
\\
(i) The $d$-quark saturation effect shifts the hyperon threshold from $\mu_B = \mu_n = M_Y$ to $2 M_Y - M_N$.
This additional $M_Y-M_N$ to the conventional $M_Y$ is the energy cost to replace a $n(udd)$ with a $Y(uds)$ at vanishing momentum, which opens the $d$-quark phase space at low momenta; 
with this opening, a neutron with the energy $\mu_n$ on the Fermi surface can decay into a $Y$,
and the resultant $Y$ saturates the $d$-quark states at low momenta.
This decay begins to occur when the energy gain $\mu_n - M_Y$ overcomes the energy cost $M_Y-M_N$.
The new threshold $\mu_n = 2 M_Y - M_N$ turns out to be very close to the threshold of the charge-neutral $S=-2$ hyperon $\Xi^0$ ($uss$), which is free from the $d$-quark saturation. 
\\
(ii) Softening associated with the appearance of $Y$ is mild.
The $d$-quark saturation constraint allows hyperons to occupy low momentum states only with small probability of $\sim 1/N_{\rm c}^3$.
Hence, hyperons must occupy higher momenta soon after their emergence.
Meanwhile,  $\Xi^0 (uss)$ hyperons, which are free from the $d$-quark saturation, can occupy states at small momenta with the probability $\sim 1$, causing significant softening, although the onset density for such doubly strange baryons is found to be large, $\sim 5$--$6\nsat$.
This density may be beyond those achievable in neutron stars.

Although the present work considerably simplifies matter with hyperons by e.g., omitting detailed descriptions of baryon-baryon interactions and structural change of baryons, we believe that the above mentioned quark saturation effects have some generic features emerging from the quark matter formation, and our statistical considerations should help to infer or constrain the trend of in-medium many-baryon forces.
We expect that convoluting the knowledge of the bare baryon-baryon interactions, being studied in various methodologies, with our statistical considerations eventually leads to the full solution for the hyperon puzzle.

This paper is structured as follows. 
In Sec.~\ref{sec:flavorless} we first review the IdylliQ model in the simplest flavor symmetric setup.
In Sec.~\ref{sec:multiflavor} we extend the IdylliQ model to the multi-flavor case.
We consider $n$-$\Lambda_0$-$\Sigma_0$ matter as $\Lambda_0$ is usually the main cause of the hyperon puzzle.
In Sec.~\ref{sec:discussion}, we add $\Xi_0 (uss)$ free of the $d$-quark saturation, and give a brief account for non-appearance of $\Delta_0 (udd) $ baryons.
In Sec.~\ref{sec:summary}, we summarize and conclude our work.

\section{Preliminaries: Flavorless Ideal Dual Quarkyonic (IdylliQ) Model}
\label{sec:flavorless}

In this section, we review the details of the ideal dual Quarkyonic (IdylliQ) model, which is necessary to understand the content of the current paper.
This is an elaboration of the work presented in Ref.~\cite{Fujimoto:2023mzy}, with some extra detail included.
The results of this section will be needed for our later analysis.

\subsection{Model}
We assume a duality relation between the momentum distribution function for quarks with a given color $f_q(q)$ and the baryon distribution $f_B(k)$ as (we denote: $\int_{k} \equiv \int \frac{d^d \bk}{(2\pi)^d} $ with $d=3$)~\cite{Kojo:2021ugu, Kojo:2021hqh}.
\begin{equation}
  \label{eq:duality}
  \hspace{-0.3cm}
  f_q (q)
  = \int_k \varphi\big(\bq - \tfrac{\bk}{N_c}\big) f_B(k)\,.
\end{equation}
Here $\varphi$ represents a single quark momentum distribution in a single baryon state. 
As quarks are confined in a spatial domain of the baryon size $\sim \Lambda_{\mathrm{QCD}}^{-1}$, $\varphi (q) $ is spread to momenta of $\sim \Lambda_{\mathrm{QCD}}$.
Adding the quark contributions from each baryon leads to quark distributions in dense matter.
Here, for symmetric nuclear matter, we include a spin-isospin degeneracy factor $g=4$.
For pure neutron matter, the factor is $g=2$.
The extension for multi-flavors and multi-baryon species will be discussed in the next section.

The normalization is $\int_q \varphi(q) = 1$.
The dual expression of the baryon number that follows from Eq.~\eqref{eq:duality} is
\begin{equation}
  \label{eq:duality_n_e}
   n_B = g\int_k f_B(k) = g \int_q f_q(q) \,.
\end{equation}
Here $f_q$ is defined for a fixed color, $f_q \equiv f_q^{R} = f_q^{G} = f_q^{B} $ with which $n_B=n_q^R=n_q^G=n_q^B$. 

For a Hamiltonian or energy density, we neglect all interactions other than that tacitly included to confine quarks into a baryon.
The energy densities are simply
\begin{equation}
    \varepsilon_{B} [f_B] = g \int_k E_B(k) f_B(k)  \,.
\end{equation}
The energy of a baryon is $E_B(k) = \sqrt{M_B^2 + k^2}$.
If we assume a simple additive quark model, one can write the baryon energy as
\begin{equation}
    E_B (k) = N_c \int_q E_q (q) \varphi \big( \bq - \bk/N_c \big)  \,,
\end{equation}
where $E_q$ is defined through this relation.
The following relation readily follows
\begin{equation}
  \varepsilon_{B} [f_B] = g \int_q E_q(q) [N_c f_q(q)] = \varepsilon_{q} [f_q]\,.
\end{equation}
This is the duality for the energy density.
The expression with $f_q$ is more useful when we examine EoS from the quark matter viewpoint.

In this work, we keep using the same $\varphi$ from low and high densities.
In reality, $\varphi$ can be modified by quark exchanges as well as structural changes in baryons.
All these effects will be considered elsewhere.

In order to determine the optimized distributions for baryons and quarks, we minimize the energy density functional with a given $n_B$ by varying $f_B$ (or $f_q$) at each momentum.
It is tempting to introduce the Lagrange multiplier to impose the fixed $n_B$ constraint, but the current problem with infinitely many variables ($f_B$ at each $k$) does not satisfy the conditions requested for the Lagrange multiplier method which demands the existence of the multiplier $\lambda$ such that $\delta \varepsilon/\delta f_B (\bk) =\lambda \delta n_B/\delta f_B(\bk)$ for all $\bk$. 
Hence we use the canonical method in which we manifestly fix $n_B$. The constraint is
\begin{equation}
0 = \int_k \delta f_B(k) \,.
\end{equation}
A variation which satisfies the constraint is
\begin{equation}
\delta f_B(\bk_1) + \delta f_B(\bk_2) = 0 \,,
\label{eq:nb_fixed_vary}
\end{equation}
which simply means we move a particle from $\bk_2$ to $\bk_1$.
The change of energy associated with such change is
\begin{align}
g^{-1}\delta \varepsilon_B
&= E_B(\bk_1) \delta f_B(\bk_1) + E_B(\bk_2) \delta f_B(\bk_2)
\notag \\
&= \big[ E_B(\bk_1) - E_B(\bk_2)\big] \delta f_B(\bk_1) \,.
\end{align}
If $E_B(\bk_1) < E_B(\bk_2)$, the energy is reduced by increasing $f_B(\bk_1)$ and reducing $f_B(\bk_2)$ at the same amount.
In particular, if we set $|\bk_2|$ to $\ksh$ which is the largest possible momentum for nonzero $f_B$, then a positive $\delta f_B(\bk)$ with $|\bk| \le \ksh$ always reduces the energy.
This describes a process in which a particle with the momentum $\bk_{\rm sh}$ decays into a particle with $\bk$.

With $|\bk_2|=\ksh$, the number-conserving variation of the energy is defined to be
\begin{align}
g^{-1} \frac{\, \delta' \varepsilon_{B} \,}{\, \delta' f_B(\bk) \,}
&\equiv
g^{-1} \bigg( 
\frac{\, \delta \varepsilon_{B} \,}{\, \delta f_B(\bk) \,}
- \frac{\, \delta \varepsilon_{B} \,}{\, \delta f_B(\bk_{\rm sh}) \,} 
\bigg)
\notag \\
&
= E_B(k) - E_B(\ksh)\,.
\end{align}
The slope is positive for $k \ge \ksh$, so $f_B(k) =0$ is favored. For $k \le \ksh$, larger values of $f_B(k)$ are favored, but whether $f_B(k)$ reaches 1 (the maximum) or not depends on the quark Pauli blocking constraint. But expressing the constraint in terms of $f_B$ is a nontrivial problem and this is the major difficulty to solve the minimization problem with the constraint.
In the IdylliQ model, we choose a specific form of $\varphi$ that allows us to invert the sum rule and express $f_B$ in terms of $f_q$.
Then the minimization problem with the constraint can be solved explicitly.

\subsection{A solvable model and its solution for the phase-space distributions}\label{sec:solution_two_flavor}

In the IdylliQ model, we choose the following distribution for $\varphi$,
\begin{equation}
  \label{eq:wf3d}
  \varphi(\bq) = \frac{2\pi^2}{\Lambda^3} \frac{e^{-q/\Lambda}}{q/\Lambda}\,,~~~~~~
  \Lhat [\varphi (\bq) ] = \frac{\, (2\pi)^2 \,}{\Lambda^2} \delta (\bq) \,
\end{equation}
where $ \Lhat = - \nabla_q^2 + \frac1{\Lambda^2}$
with $\Lambda $ being some scale parameter characterizing the size of baryons.
Applying this operator to Eq.~\eqref{eq:duality}, we find a local relation between $f_B$ and $f_q$ ($d=3)$,
\begin{equation}
  \label{eq:derivrel}
  f_B(N_c q) =  \frac{\Lambda^2}{N_c^d} \, \Lhat \big[ f_q(q) \big] \,.
\end{equation}

The candidates of local solutions for $f_q$ are the boundary values $f_q=1$ and those dual to $f_B=0$ and $1$.
The $f_q$ dual to $f_B=0$ and $1$ are determined as follows.
The most general solution for $f_B = 0$ is
\begin{align}
    f_q^{f_B=0} (q) = c_+ y_+ (q) + c_- y_- (q)\,,
\end{align}
where $y_\pm (q) = e^{\pm q/\Lambda}/q$ satisfying  $\Lhat [y_\pm] = 0$.
Meanwhile $f_B (N_c q) = 1$ can be obtained as 
\begin{align}
    f_q^{f_B=1} (q) = N_c^d + d_+ y_+ (q) + d_- y_- (q)\,.
\end{align}
The range of constants $c_\pm$ and $d_\pm$ are constrained by the condition $0 \le f_q(q) \le 1$.

We now put the candidates of solutions together.
Before quark states are saturated, however, we do not have to consider the quark Pauli blocking constraint and one can readily determine the baryon distribution as 
\begin{equation}
f_B (k) = \Theta( \ksh - k ) \,,
\end{equation}
and we can substitute it into the sum rule.
The solution for $f_q(q)$ takes the form
\begin{equation}
f_q (q) = f_q^{f_B=1} (q) \Theta (\qsh - q ) + f_q^{f_B=0} (q) \Theta (q - \qsh ) \,,
\label{eq:fq_before}
\end{equation}
with coefficients $c_\pm$ and $d_\pm$ determined from the sum rule. 
We will relate $\ksh$ and $\qsh$ shortly, and it will turn out $\ksh = N_c \qsh$.
The expression is valid for sufficiently small $\ksh$ with which $f_q(0)\le 1$. Beyond some value of $\ksh$, the extrapolation of Eq.~\eqref{eq:fq_before} for large $\ksh$ would violate the condition $f_q(0) \le 1$.
The saturation at $q=0$ first occurs at $ k_{\rm F} \approx \sqrt{2/N_c} \Lambda$ or $ n_B /\nsat = 2.58 \times \big( \Lambda /0.4\,\text{GeV} \big)^3$ for $N_c =3$.
Note that this implies the saturation density is parametrically small compared to the QCD scale $\Lambda^3$.

When quark states are filled at low momenta the solution should take the form
\begin{align}
    f_q(q) = \Theta (\qbu - q )
    &+ f_q^{f_B=1} (q) \Theta (\qsh - q )\Theta (q - \qbu ) 
    \notag \\
    &+ f_q^{f_B=0} (q) \Theta (q - \qsh ) \,.
  \label{eq:fQpostsat}
\end{align}
For $q \le \qbu$ the quark states are saturated.
Beyond $q \ge \qbu$ the quark distribution is dual to the saturated baryon distribution; here the baryons become free from the Pauli blocking constraint and hence $f_B$ can take the maximal value 1.
For $q \ge \qsh$, the quark distribution is dual to the baryon distribution where $f_B$ is zero.
Applying $\hat{L}$, we find
\begin{align}
\frac{\Lambda^2}{N_c^d} \, \Lhat \big[ f_q(q) \big] 
= \frac{1}{\, N_c^d \,} \Theta (\qbu - q)
&+ \Theta (\qsh - q )\Theta (q - \qbu)
\notag \\
&+ \frac{\Lambda^2}{N_c^d} R \,,
\label{eq:fBpostsat_}
\end{align}
where $R$ comes from the application of derivatives on step functions,
\begin{align}
R = 
& \big[ 1 - f_q^{f_B=1} (\qbu) \big] \delta' (\qbu - q )
\notag \\
&+ \big[ f_q^{f_B=1} (\qsh)-f_q^{f_B=0} (\qsh) \big] \delta' (\qsh - q )
\notag \\
&
- \nabla f_q^{f_B=1} (\qbu) \delta (q - \qbu )
\notag \\
& + \nabla \big[ f_q^{f_B=1}(\qsh) - f_q^{f_B=0}(\qsh) \big] \delta (q - \qsh )  
\,.
\end{align}
The function $R$ would contain the $\delta$- and $\delta'$-functions which violate the condition $0\le f_B \le 1$. This problem can be avoided by demanding $f_q$ and $d f_q/d q$ to be continuous at each junction point. The resulting $R$ vanishes and we get ($k=N_c q, \kbu=N_c \qbu, \ksh = N_c \qsh$)
\begin{equation}
f_B (k) = \frac{1}{\, N_c^d \,} \Theta (\kbu - k)
+ \Theta (\ksh - k )\Theta (k - \kbu) \,.
\label{eq:fBpostsat}
\end{equation}
The rest is the determination of $\kbu$ that follows from the continuity conditions of $f_q$. This in turn requires the determination of $c_\pm$ and $d_\pm$. One can set $c_+=0$ to satisfy $f_q \rightarrow 0$ for $q\rightarrow \infty$.
So four conditions are sufficient to determine $\kbu$, $c_-$, and $d_\pm$.

With four conditions from two junction points, one can express $c_-$, $d_\pm$,
and $\Delta_{q} = \qsh - \qbu$ as functions of $\qsh$.
The conditions are
\begin{align}
\label{eq:fb1_con}
    \begin{split}
    1 &= f_q^{f_B=1}(\qbu)\,,\\
    0 &= \frac{df_q^{f_B = 1}(\qbu)}{dq}\,, \\
    f_q^{f_B=1}(\qsh) &= f_q^{f_B=0}(\qsh)\,, \\
    \frac{df_q^{f_B=1}(\qsh)}{dq} &= \frac{df_q^{f_B=0}(\qsh)}{dq}\,.
    \end{split}
\end{align}
the four coefficients are entirely determined by these four matching equations.

After lengthy calculations, the matching coefficients and $\Delta_q$ can be found as functions of $\qbu$ and $\qsh$.
\begin{equation}
    N_c^3 = \frac{\calX_{N_c^3}}{\calY}\,,\qquad
    d_{\pm} = \frac{\calX_{d_\pm}}{\calY}\,,\qquad
    c_- = \frac{\calX_{c_-}}{\calY}\,,
\end{equation}
where the numerators are
\begin{align}
  \calX_{N_c^3} &= -[y_{+}'(\qsh) y_{-}(\qsh) - y_{+}(\qsh) y_{-}'(\qsh)]y_{-}'(\qbu)\,, 
  \notag \\
  \calX_{d_+} &= - y_{-}'(\qbu) y_{-}'(\qsh)\,,
    \notag \\
  \calX_{d_-} &= y_{+}'(\qbu) y_{-}'(\qsh)\,,
    \notag \\
  \calX_{c_-} &= y_{+}'(\qbu) y_{-}'(\qsh) - y_{-}'(\qbu) y_{+}'(\qsh)\,,
\end{align}
and the common denominator is
\begin{align}
  \calY &= [y_{+}'(\qbu) y_{-}(\qbu) - y_{+}(\qbu) y_{-}'(\qbu)]y_{-}'(\qsh) 
    \notag \\
  & - [y_{+}'(\qsh) y_{-}(\qsh) - y_{+}(\qsh) y_{-}'(\qsh)]y_{-}'(\qbu)\,.
\end{align}
Here we display only the equation to determine $\Delta_{q}$, which is obtained from the relation $N_c^3 = \calX_{N_c^3} / \calY$;
it is needed for the computations of EoS.
The equation to be solved is
\begin{equation}
  \label{eq:hkFB}
   \frac{\Lambda + \qbu}{\Lambda + \qbu - (\Lambda + \qsh) e^{-\Delta_{q} / \Lambda}} = N_c^3 \,.
\end{equation}
Using $\qbu$ and $\qsh$ determined above,
the baryon distribution in \eqref{eq:fBpostsat} leads to the EoS
\begin{align}
    n_B
    &= 4\int_{\ksh - \Delta}^{\ksh} \frac{d^3 \bk}{(2\pi)^3}
    + \frac{4}{N_c^3} \int_0^{\ksh - \Delta} \frac{d^3 \bk}{(2\pi)^3}\,,\\
    \varepsilon
    &= 4\int_{\ksh - \Delta}^{\ksh} \frac{d^3 \bk}{(2\pi)^3} \sqrt{k^2 + M_N} 
    \notag \\
    &~~~ + \frac{4}{N_c^3} \int_0^{\ksh - \Delta} \frac{d^3 \bk}{(2\pi)^3} \sqrt{k^2 + M_N}\,.
\end{align}
This is the flavor symmetric, two-flavor EoS in the IdylliQ model.

The EoS in the IdylliQ model is similar (but not quite identical) to the one proposed in McLerran-Reddy (MR) model~\cite{McLerran:2018hbz}. 
In the second integral in $\varepsilon$, we make the change of the variable $\bk \to \bq = \bk / N_c$
and assume the relation $M_N \rightarrow  N_c M_q $, finding the form in Ref.~\cite{McLerran:2018hbz},
\begin{align}
    n_B^{\rm MR} 
    &= \frac{3}{2\pi^2} \left[\ksh^3 - (\ksh - \Delta)^3 + \frac{(\ksh - \Delta)^3}{N_c^3}\right]\,,
        \label{eq:MR_n}
    \\
    \varepsilon^{\rm MR}
    &= 4\int_{\ksh - \Delta}^{\ksh} \frac{d^3 \bk}{(2\pi)^3} \sqrt{k^2 + M_N^2} 
    \notag \\
    &~~~ + 4 N_c \int_0^{\frac{\ksh - \Delta}{N_c}} \frac{d^3 \bq}{(2\pi)^3} \sqrt{q^2 + M_q^2}
    \label{eq:MR_e}
    \,.
\end{align}
The only difference in EoS between the IdylliQ and MR models
lies in the explicit form of the quark kinetic energy.
The two EoS become completely identical if one replaces
$ \sqrt{q^2 + M_q^2}$ with $E_q^{\rm IdylliQ} (\bq)$ {\it defined} through the relation
\begin{equation}
\sqrt{k^2 + M_N^2}
= N_c \int_q E_q^{\rm IdylliQ} (\bq) \varphi\big(\bq - \tfrac{\bk}{N_c}\big) \,.
\end{equation}
Applying the operator $\Lhat$ acting on the variable $\vk$, 
one can explicitly derive the form of $E_q^{\rm IdylliQ}  (q)$ which reflects 
that quarks are confined into a baryon.

\subsection{Comparison with the previous model of Quarkyonic Matter}
\label{sec:previousmodel}

\begin{figure}
    \centering
    \includegraphics[width=0.96\linewidth]{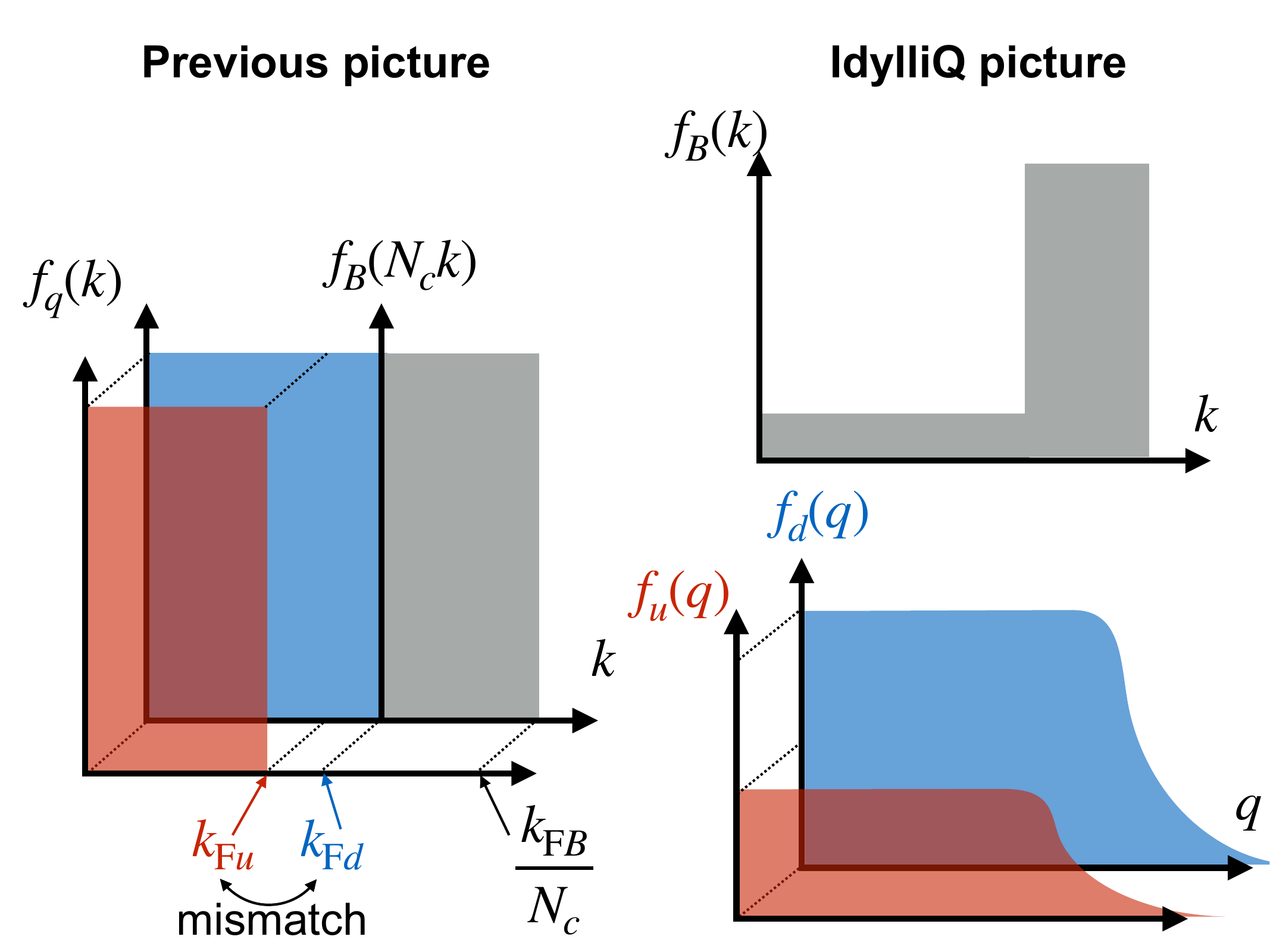}
    \caption{Comparison of the previous model of Quarkyonic matter and IdylliQ model.  
    In the previous modeling,
    there is a discrepancy between the Fermi momenta of $u$ and $d$ quarks, so the $u$-quark distribution has a mismatch with the nucleon distribution. Meanwhile, in the IdylliQ picture, this mismatch is naturally resolved by considering the half-occupied distribution for the $u$ quark that is dual to the nucleonic description.}
    \label{fig:comparison}
\end{figure}

The extension of the above-mentioned flavor symmetric model \cite{McLerran:2018hbz} to the flavor asymmetric one is not straightforward.
The previous attempts extended the form \eqref{eq:MR_e} 
by generalizing the flavor symmetric quark part into the flavor asymmetric ones.
But without including the physics that quarks are bound to baryons,
such extension leaves a mismatch between baryon and quark momentum distributions.
For instance, in pure neutron matter,
typical attempts to fill $u$- and $d$-quark states with the probabilities $1$
leads to the $u$- and $d$-quark Fermi sea at different size,
see the leftmost panel in Fig.~\ref{fig:comparison}.
Then, picking $u$- and $d$-quarks near the Fermi surfaces result in a baryon having $u$- and $d$-quark momentum distributions
considerably deformed from those in ordinary baryons.
In domains not very far from nuclear matter densities, however,
we expect that highly deformed baryons are energetically disfavored.
In this work, we instead assume that confining forces remain strong enough to 
maintain the overall structure of baryons,
and as a result the quarks have more general momentum distribution than in a simple Fermi sea,
as shown in the rightmost panels in Fig.~\ref{fig:comparison}.

\section{IdylliQ model with multiple flavors}
\label{sec:multiflavor}

In this section, we extend the IdylliQ model reviewed in the previous section to the case with multiple flavors.
The matter composition can be determined by minimizing the energy density for a fixed baryon density.
Equivalently, one can determine the composition by satisfying the chemical equilibrium condition.

\subsection{Simplified setup}
Here we consider adding the hyperons $\Lambda^0$ and $\Sigma^0$ on top of pure neutron matter.
As already mentioned, we assume the mass of $\Lambda^0$ and $\Sigma^0$ are degenerate, i.e., $M_Y \equiv M_{\Lambda^0} = M_{\Sigma^0}$, and denote the degeneracy factor as $d_Y = 2$.
Throughout this paper, we use the value $M_N \approx 0.94\,\text{GeV}$ and $M_Y \approx 1.12\,\text{GeV}$.
In this simple setup, we have only two types of variables, $f_N(k)$ and $f_Y (k) \equiv f_{\Lambda^0}(k) = f_{\Sigma^0}(k)$.

The energy density described in terms of the momentum distribution of hadrons $f_i(k)$ $(i = n, Y)$ is
\begin{align}
    \varepsilon = g \int_{\bk} \left[E_N (k) f_n(k) + d_Y E_Y(k) f_Y(k) \right]\,,
\end{align}
A factor $g = 2$ in front of the integral is the degeneracy of spins, and we use the ideal dispersion relation $E_i(k) = \sqrt{k^2 + M_i}$ as before.
The number density of neutrons and hyperons are
\begin{align}
    n_n = g\int_{\bk} f_n(k)\,,\quad 
    n_Y = g\int_{\bk} d_Y f_Y(k)\,.
\end{align}
The charge neutrality $n_Q=0$ is fulfilled trivially as we only consider the charge-neutral baryons.

Now, consider the momentum distribution of quarks: $f_q(q)$ $(q = u, d, s)$.
The confinement relation that sets the duality transformation between $f_B(k)$ and $f_q(q)$ is
\begin{align}
    f_q(q) = \sum_{i=n,\Sigma^0,\Lambda^0} \int_{\bk} \varphi \left(\bq - \frac{\bk}{N_c}\right) B_{q}^i f_i (k)\,,
\end{align}
where $B_q^i$ is the baryon number carried by a quark of flavor 
$q$ inside the baryon species $i$.
Explicitly,
\begin{align}
   \hspace{-0.2cm} f_d(q) &= \int_{\bk} \varphi \left(\bq - \frac{\bk}{N_c}\right) \left[B_{d}^{n} f_n(k) + d_Y B_{d}^{Y} f_Y(k) \right]\,, \label{eq:d-sat}
   \\
   \hspace{-0.2cm} f_u(q) &= \int_{\bk} \varphi \left(\bq - \frac{\bk}{N_c}\right) \left[B_{u}^{n} f_n(k) + d_Y B_{u}^{Y} f_Y(k)  \right]\,,\\
   \hspace{-0.2cm} f_s(q) &= \int_{\bk} \varphi \left(\bq - \frac{\bk}{N_c}\right) \left[d_Y B_{s}^{Y} f_Y(k)\right]\,,
\end{align}
where $B_d^n = 2/3$ and $B_{d}^{Y} = 1/3$ are the baryon number carried by constituent $d$ quarks in neutrons and the average baryon number carried by constituent $d$ quarks in hyperons.
Also, $B_u^n = 1/3$ and $B_u^Y = 1/3$ for $u$ quarks in baryons, and $B_{s}^{Y} = 1/3$ for $s$ quarks in baryons.

We consider the case in which the $d$ quark state saturates.
The low-$k$ regions of $f_n(k)$ and $f_Y(k)$ are subject to the saturation of $d$-quark states $f_d(q)$ at $q\le \qbu$.
Applying $\Lhat$ to Eq.~\eqref{eq:d-sat} with $f_d=1$,
we find the constraint 
\begin{align}
    B_{d}^{n} f_n^{\rm bulk}(k) + d_Y B_{d}^{Y} f_Y^{\rm bulk}(k) = \frac{1}{N_c^3}\,,
    \label{eq:satky}
\end{align}
at $k\le \kbu = N_c \qbu$.
With the constraint, we optimize the $f_n$ and $f_Y$ to minimize the energy. 

We consider the energy minimization with $n_B$ fixed. As before we discuss how the particles with the maximal energy transform into particles with less energy. 
We assume that the shell part is saturated by neutrons while the bulk may be a mixture of neutrons and hyperons
\begin{align}
    \label{eq:fn}
    f_n(k) &= f_n^{\rm bulk} (k) \Theta(\kbu - k) + \Theta(k - \kbu) \Theta(\ksh - k)\,,\\
    f_Y(k) &= f_Y^{\rm bulk} (k) \Theta(\kbu - k) \,,
    \label{eq:fY}
\end{align}
where $f_n^{\rm bulk} (k)$ and $ f_Y^{\rm bulk} (k)$ satisfy the $d$-quark saturation condition Eq.~\eqref{eq:satky}.
To optimize the composition in the bulk, we consider the conversion $n (k)\rightarrow Y (k)$ at $k < \kbu$.
Here we have introduced the notation $B(k)$ which represents a baryon $B$ with the momentum $k$.

\subsection{Determination of the bulk composition}

As done in Sec.~\ref{sec:flavorless} for isospin symmetric matter, we vary $f_n$ and $f_Y$ in the bulk to minimize the energy functional for a fixed $n_B$.
Unlike the previous case where the application of $\Lhat$ to $f_q$ directly determines $f_N$,
here we further must determine the composition through the energy consideration. 

The saturation condition imposes important constraints on the particle conversion processes.
Even when $\mu_n$ exceeds $M_Y$, the weak decay $n(\ksh) \rightarrow Y(k=0)$ does not necessarily occur
if the $d$-quark states are saturated.
An extra process to open $d$-states in the bulk is needed for such decays, demanding an extra energy cost.

When one neutron converts into one hyperon, $n(k) \rightarrow Y(k)$, this opens the phase space for another hyperon to come to fill the state $Y(k)$, 
since neutrons contain two $d$-quarks while each of $\Sigma^0, \Lambda^0$ contains only single $d$-quark.
This extra hyperon can be brought by the conversion of neutrons in the shell into hyperons in the bulk, $n(\ksh) \rightarrow Y(k)$,
and this decay reduces the energy of the system.
Through these sequences of the processes, not only the bulk but also the shell structure in momentum space are reorganized.

One can examine the energy gain and cost for each process while keeping $n_B$ fixed.
This can be manifestly done by varying $f_n$ and $f_Y$ as in Eq.~\eqref{eq:nb_fixed_vary}.
The calculations are rather lengthy and we postpone the step-by-step calculations to the Appendix~\ref{sec:one-by-one}.
Below, we show quicker calculations by focusing on the chemical equilibrium conditions. 

The key observation to skip complications of number-conserving energy variation is that, when we reach the ground state, the conversion processes reach the equilibrium; knowing this fact allows us to impose the $\beta$-equilibrium condition from the very beginning for an efficient search for the ground state.
We prepare several candidates of the solution (parametrization of $f_n$ and $f_Y$), 
demand the $\beta$-equilibrium condition to fix the parameters in each candidate, and then single out the minimum solution at the same $n_B$.

For simplicity, we consider a sufficiently low density close to the hyperon thresholds.
We can safely assume that those hyperons appear at low momenta, with the occupation probability being constant as we justify shortly.
Therefore, we take the following ansatz for the solution:
\begin{align}
    \label{eq:fn}
    f_n(k) &= \frac{h_n}{N_c^3} \Theta(k_Y - k)\Theta(k) 
    \notag \\
    &~~~ + \frac{1}{B_{d}^{n} N_c^3} \Theta(k - k_Y) \Theta(\kbu - k) 
    \notag \\
    &~~~ + \Theta(k - \kbu) \Theta(\ksh - k)\,,\\
    f_Y(k) &= \frac{h_Y}{ N_c^3} \Theta(k_Y - k)\Theta(k)\,.
    \label{eq:fY}
\end{align}
From Eq.~\eqref{eq:satky}, the parameters $h_n$ and $h_Y$ satisfies
\begin{equation}
    B_{d}^{n} h_n + d_Y B_{d}^{Y} h_Y = 1\,.
    \label{eq:satky2}
\end{equation}
This implies that $0 \leq h_n \leq 1 / B_{d}^{n}$ and $0 \leq h_Y \leq 1/(d_Y B_{d}^{Y})$.
Using these distributions and eliminating $h_n$ by using Eq.~\eqref{eq:satky2} in the above equations, the thermodynamic quantities are
\begin{align}
    \label{eq:nn}
    \frac{\, n_n \,}{g} &= - d_Y 
    \frac{h_Y }{ N_c^3} \frac{B_{d}^{Y}}{B_d^{n} }
    \int_0^{k_Y} \!\!\!\!\!\! dk\, D(k) 
    \notag \\
    &~~~
    - \left(1 - \frac{1}{B_{d}^{n} N_c^3}\right) \int_0^{\kbu} \!\!\!\!\!\! dk\, D(k) 
    + \int_0^{\ksh} \!\!\!\!\!\! dk\, D(k)\,,\\
    \label{eq:nY}
    \frac{\, n_Y \,}{g} &= d_Y  \frac{ h_Y }{ N_c^3} \int_0^{k_Y} \!\!\!\!\!\! dk\, D(k)\,,
    \\
    \frac{\, \varepsilon \,}{g} &=  d_Y \frac{ h_Y }{N_c^3} \int_0^{k_Y} \!\!\!\!\!\! dk\, D(k) 
    \left[ E_Y(k) 
    - \frac{ B_{d}^{Y} }{B_{d}^{n}} E_N(k) \right] 
    \notag \\
    &~~~ 
    -  \left(1 - \frac{1}{B_{d}^{n} N_c^3}\right) \int_0^{\kbu} \!\!\!\!\!\! dk\, D(k) E_N(k) 
    \notag \\
    &~~~
    + \int_0^{\ksh} \!\!\!\!\!\! dk\, D(k) E_N(k)\,,
    \label{eq:enY}
\end{align}
where the density of states is given by $D(k) = k^2 / (2\pi^2)$.

Now we have four parameters, $h_Y, k_Y, \kbu$, and $\ksh$ to be determined.
As we mentioned, we impose the $\beta$-equilibrium condition.
The (candidates of) chemical potentials $\mu_n$ and $\mu_Y$ are given by
\begin{align}
    \mu_n = \left(\frac{\partial \varepsilon}{\partial n_n}\right)_{n_Y}\,,\quad
    \mu_Y = \left(\frac{\partial \varepsilon}{\partial n_Y}\right)_{n_n}\,.
\end{align}
The chemical potential for neutrons $\mu_n$ and $\mu_Y$ can be calculated by using the method of the Jacobian.
The details of calculations are given in Appendix~\ref{app:Jacobi}.
The neutron chemical potential is found to be
\begin{align}
    \mu_n &= \frac{
    - D E_N \big|_{\kbu} \left(1 - \frac{1}{B_{d}^{n} N_c^3}\right) \frac{\partial \kbu}{\partial \ksh}
    + D E_N \big|_{\ksh} }
    {- D(\kbu)\left(1 - \frac{1}{B_{d}^{n} N_c^3}\right) \frac{\partial \kbu}{\partial \ksh} + D(\ksh) }\,,
\end{align}
which is independent of hyperon parameters $k_Y$ and $h_Y$; it is essentially a function of $\ksh$ as $\kbu$ is fixed through the condition $f_n(k) = 1$ for $\kbu \le k \le \ksh$, in the similar way as Eq.~\eqref{eq:fb1_con}.
Meanwhile, the hyperon chemical potential is
\begin{align}
    \mu_Y
    &= E_Y(k_Y) - \frac {B_{d}^{Y} }{ B_{d}^{n} }E_N(k_Y)
     + \frac{ B_{d}^{Y} }{ B_{d}^{n}} \mu_n\notag\\
     &= E_Y(k_Y) - \frac12 E_N(k_Y) + \frac12 \mu_n\,,
     \label{eq:muY}
\end{align}
where in the last line, we used $B_{d}^{Y} = 1/3$ and $B_{d}^{n} = 2/3$.

Recalling that $\mu_n$ is a function of $\ksh$, the chemical equilibrium condition $\mu_n = \mu_Y = \mu_B$ turns into the relation between $\ksh$ and $k_Y$ for a given $h_Y$, 
which reads
\begin{equation}
\mu_B (\ksh) = 2E_Y (k_Y) - E_N (k_Y) \,.
\label{eq:ne_con}
\end{equation}
This is the necessary condition for the ground state. As we have derived the condition at a given $h_Y$, the parameters $\ksh$ and $k_Y$ should be regarded as functions of $h_Y$.

Finally, we determine $h_Y$ by choosing the lowest energy solution out of candidates satisfying Eq.~\eqref{eq:ne_con}.
The variation of $h_n$ while holding $n_B$ fixed yields
\begin{align}
    \left(\frac{\partial \varepsilon}{\partial h_Y}\right)_{n_B} 
    \!\! = \frac{\, g d_Y  \,}{N_c^3} \! \int_0^{k_Y} \!\!\!\!\! dk D(k) \left[ E_Y(k) - \frac{\, E_N(k)  + \mu_B \,}{2}
  \right]\,.
\end{align}
We note that the integrand is a monotonically increasing function of $k$,
\begin{equation}
\frac{\partial}{\partial k} \big[ E_Y(k) - \frac{1}{2} E_N(k) \big] 
= \bigg( \frac{1}{E_Y(k)} - \frac{1}{ 2E_N(k)} \bigg) k > 0 \,,
\end{equation}
provided that $E_Y(k) < 2E_N(k)$. This condition holds for $M_Y < 2M_N$, valid for realistic baryon masses.
Now we found that the integral is always negative for $k<k_Y$ and vanishes at $k=k_Y$, see Eq.~\eqref{eq:ne_con}.
Therefore $(\partial \varepsilon/\partial h_Y)_{n_B} < 0$, favoring a larger $h_Y$.
Consequently, $h_Y$ takes the largest value possible; the domain of $k<k_Y$ is fully saturated by hyperons.

\begin{figure}
    \centering
    \includegraphics[width=0.96\columnwidth]{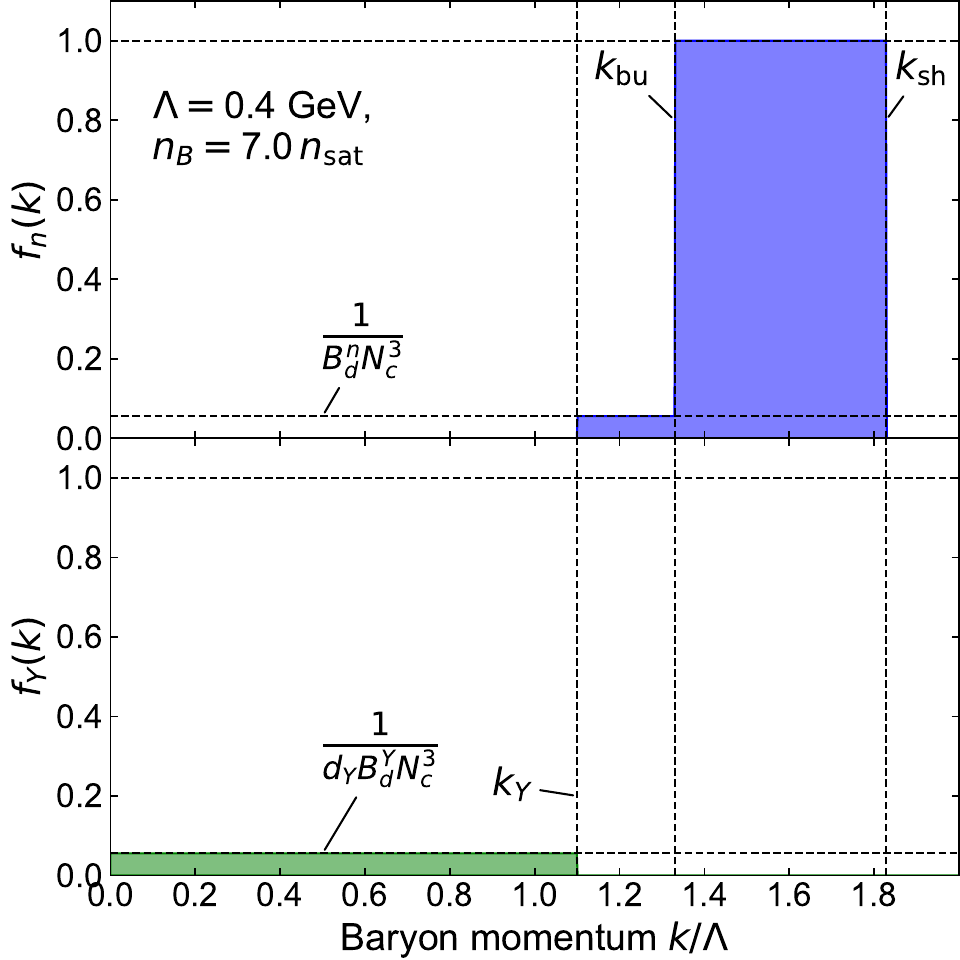}\hfill
    \includegraphics[width=0.96\columnwidth]{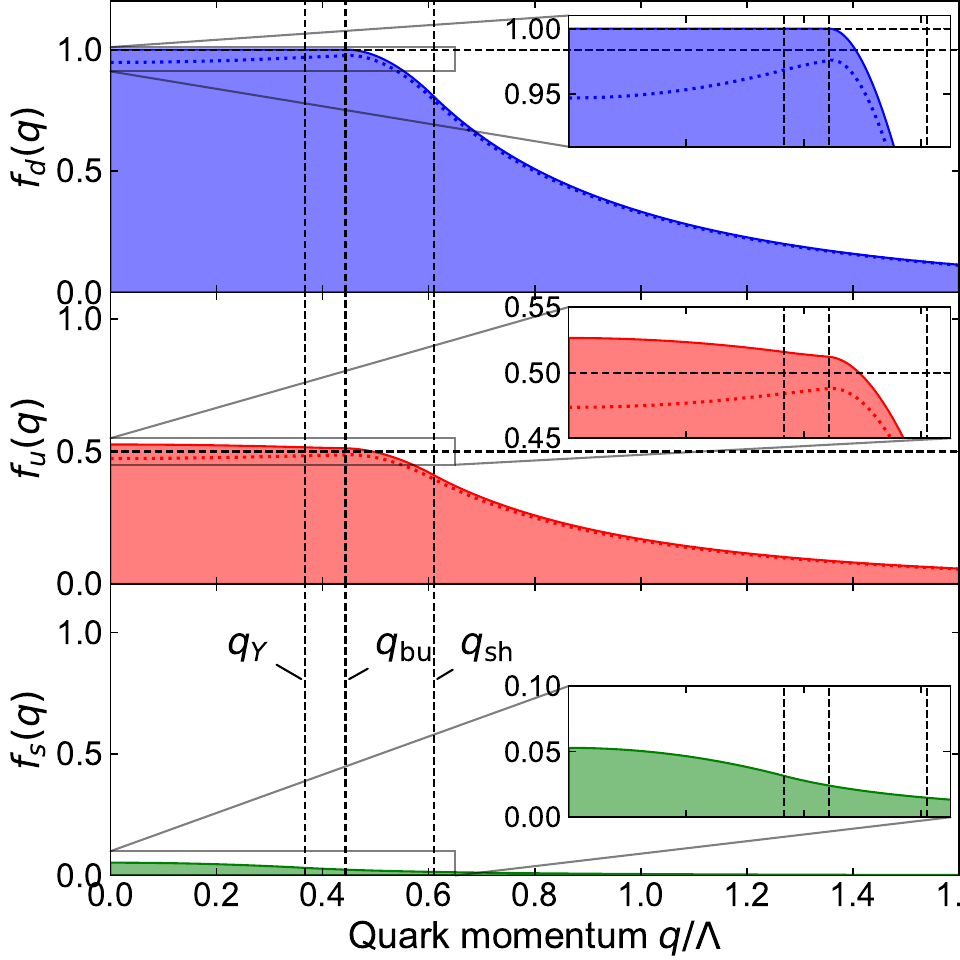}
    \caption{Phase-space density distribution for baryons (left) and quarks (right).
    We use $M_N = 0.94\,\text{GeV}$ and $M_Y = 1.12\,\text{GeV}$ ($\approx M_\Lambda < M_{\Sigma_0} \approx 1.19\,\text{GeV}$). In the left panel, $k_{\mathrm{F}n, \mathrm{ideal}}$ and $k_{\mathrm{F}Y, \mathrm{ideal}}$ refer to the Fermi momentum of $n$ and $Y$ in the ideal Fermi gas picture without quark saturation effect taken into account, respectively.
    }
    \label{fig:phasespace}
\end{figure}

Now our solution takes the form 
\begin{align}
    \label{eq:fn_opt}
    f_n(k) &=  \frac{1}{B_{d}^{n} N_c^3} \Theta(k - k_Y) \Theta(\kbu - k) \
    \notag \\
    &~~~ + \Theta(k - \kbu) \Theta(\ksh - k)\,,\\
    f_Y(k) &= \frac{ 1 }{ d_Y B_d^Y N_c^3} \Theta(k_Y - k)\,.
    \label{eq:fY_opt}
\end{align}
with Eq.~\eqref{eq:ne_con} for $h_Y = 1/(d_Y n_d^Y)$,
\begin{equation}
\mu_B = 2E_Y (k_Y) - E_N (k_Y) \,.
\label{eq:chemi}
\end{equation}
The solution is valid for sufficiently large $n_B$.
We recall that the RHS of Eq.~\eqref{eq:chemi} is an increasing function of $k_Y$ so that $\mu_B$ at the onset of hyperons is
\begin{equation}
\mu_B^{\rm onset} = 2M_Y - M_N \,,
\end{equation}
below which it is not possible to satisfy the chemical equilibrium; neutrons completely dominate the system.
A snapshot of the phase space distributions for baryons and quarks is shown in Fig.~\ref{fig:phasespace}.

\subsection{Equation of state}

In Fig.~\ref{fig:eos04}, we show the EoS~\eqref{eq:nn}-\eqref{eq:enY}.
The explicit form of the EoS is given in Appendix~\ref{app:eos}.
The onset of hyperons is shown with dash-dotted lines in each figure.

There are two remarkable consequences of the quark saturation:

First, hyperons emerge in the system at density $\sim (5$--$6)n_{\rm sat}$, significantly higher than the usual estimate, ($2$--$3)n_{\rm sat}$.
We note that the typical core density of two-solar mass neutron stars is about $\sim 5\nsat$; if our estimate on the hyperon onset is valid, the high mass is established before hyperons soften the EoS.
Second, even after the appearance of hyperons, the softening is mild as the contribution of the hyperons is suppressed by a factor $1/N_c^3$.
Since hyperons cannot largely occupy the phase space at low energy, increasing the hyperon density makes hyperons energetic more quickly than theories without the saturation;
the relativistic regime for hyperons is reached at lower density.

We emphasize that all these dramatic effects arise solely from the statistical considerations; no detailed discussions of interactions are used.
We expect that the baryon-baryon interactions, which seem to be repulsive in most flavor channels, further stiffen the EoS.

Softening may be more significant when we add $\Xi^0$, which is a heavier baryon not subject to $d$-quark saturation condition, and lies slightly above the threshold of hyperons with $S=-1$.
This issue will be explained in the next section.

\begin{figure}
    \centering
    \includegraphics[width=0.96\columnwidth]{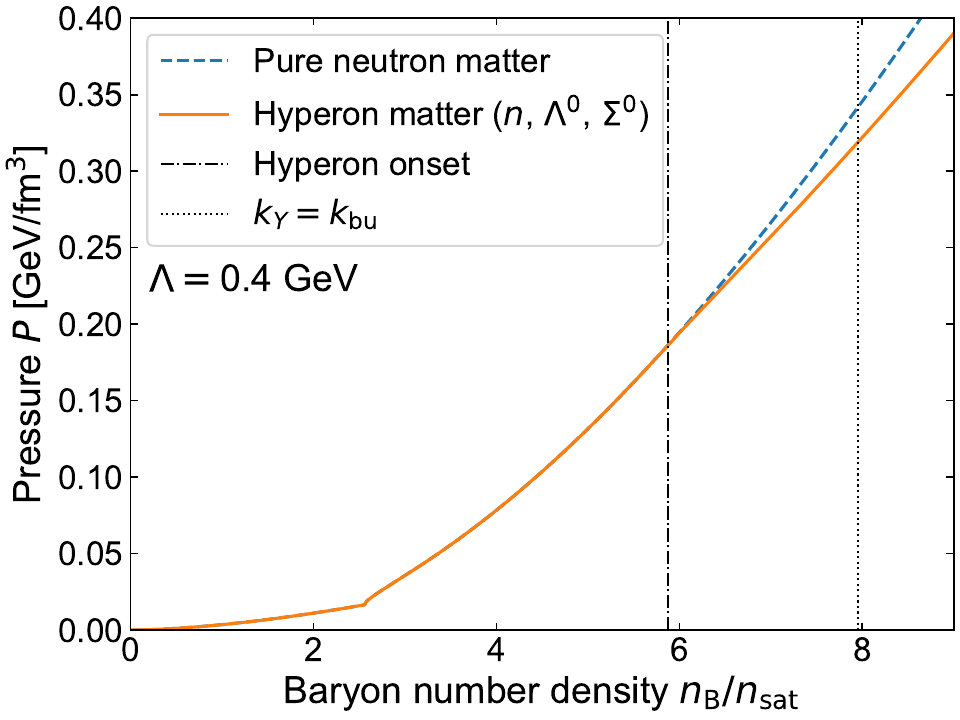} \hfill
    \includegraphics[width=0.96\columnwidth]{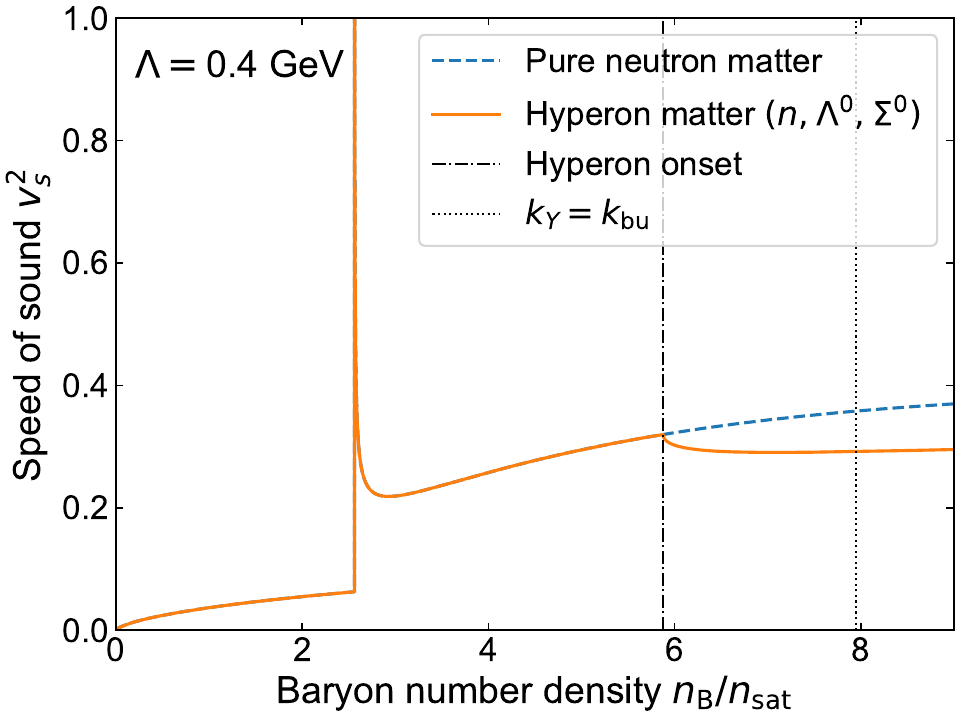}
    \caption{Comparison of the equations of state between the pure neutron matter and the hyperon matter at $\Lambda = 0.4$ GeV. Left: Pressure as a function of $n_B$.  Right: Speed of sound as a function of $n_B$.}
    \label{fig:eos04}
\end{figure}

\section{Discussion}
\label{sec:discussion}

In this section, we discuss several points regarding the solution we have given in the previous section.
We consider adding heavier baryons.
In particular, we consider $\Xi^0$, which is a hyperon with strangeness $S=-2$, and $\Delta^0$, which is a baryon without strangeness in it whose mass lies between $\Sigma^0$ and $\Xi^0$.
We will see that the former has the threshold very close to those of $S=-1$ hyperons while the latter turns out to be energetically disfavored and thus does not appear in the current setup.
Then, we mention the applicable limit of our current solution, and close with the possible future direction.

\subsection{Including heavier baryons: the case of $\Xi^0$}
\label{sec:Xi0}

Now we consider including the heavier baryons.
Among the charge-neutral hyperons, the quark content of $\Xi^0$ is $uss$, so the threshold is never affected by the $d$-quark saturation effect.
In additive quark models, the threshold is estimated to be
\begin{equation}
    (\mu_{B})_{\rm onset}^{\Xi^0} = M_{\Xi^0} ~\approx~ 2M_s + M_u \,,
\end{equation}
where $M_{\Xi^0}$ is the mass of $\Xi^0$
which is close to the onset of $\Lambda^0$ and $\Sigma^0$,
\begin{equation}
(\mu_{B})_{\rm onset}^{\Lambda^0,\Sigma^0}
= 2M_Y - M_N  
~\approx~ 2M_s + M_u\,.
\end{equation}
Therefore, near the threshold of hyperons with $S=-1$,  $\Xi^0$ also enters the system.
Below, for simplicity, we approximate $M_{\Xi^0}$ to $2M_Y - M_N$ and discuss the effect of $\Xi^0$ to the EoS.

\begin{figure}
    \centering
    \includegraphics[width=0.96\columnwidth]{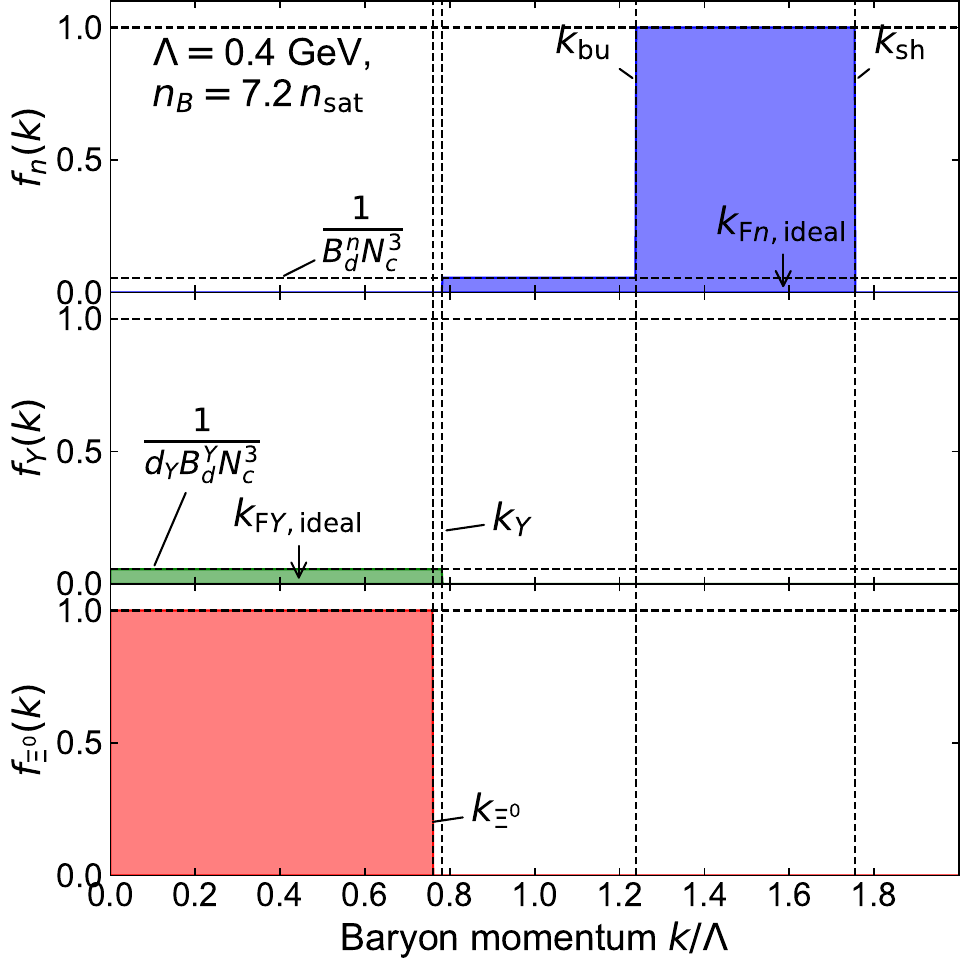}\hfill
    \includegraphics[width=0.96\columnwidth]{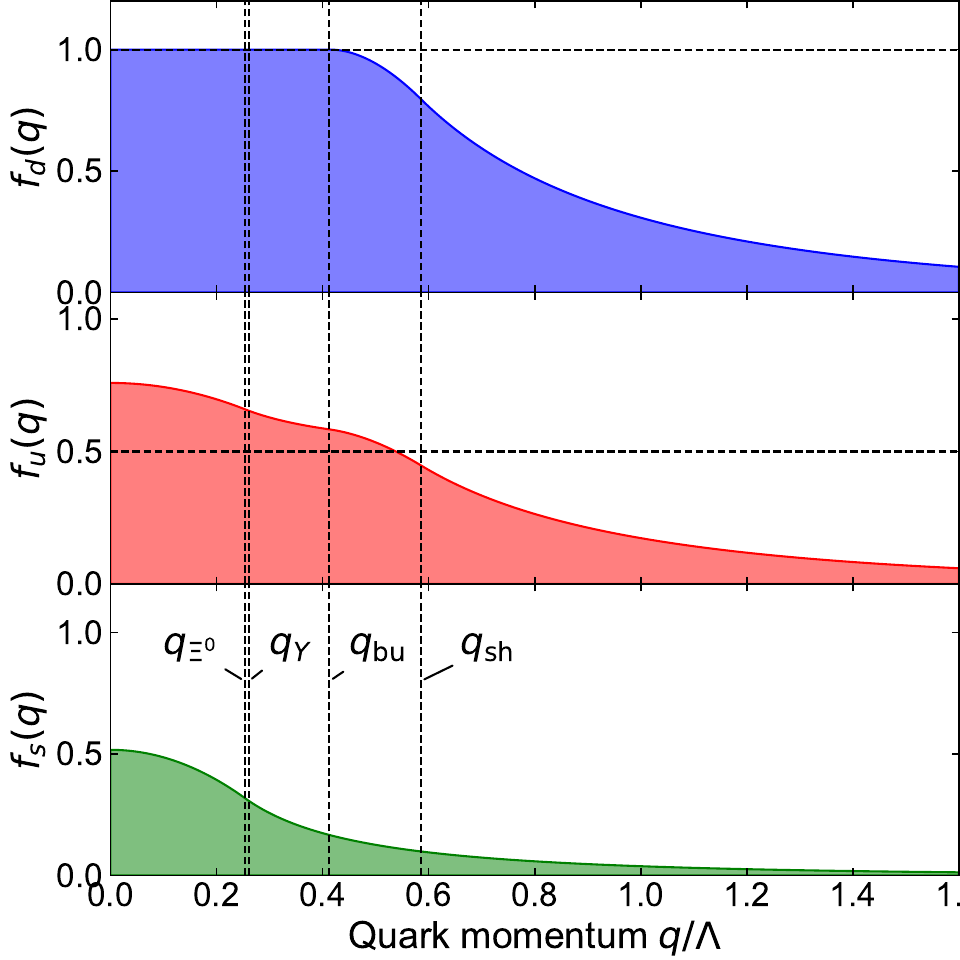}
    \caption{Phase-space density distribution for baryons (left) and quarks (right).
    In the left panel, $k_{\mathrm{F}n, \mathrm{ideal}}$ and $k_{\mathrm{F}Y, \mathrm{ideal}}$ refer to the Fermi momentum of $n$ and $Y$ in the ideal Fermi gas picture without quark saturation effect taken into account, respectively. Note that in the ideal Fermi gas picture, the $\Xi^0$ threshold has not reached yet.}
    \label{fig:phasespace_xi0}
\end{figure}

As $\Xi^0$ is never affected by the $d$-quark saturation condition, momentum distribution is given by
\begin{equation}
    f_{\Xi^0}(k) = \Theta(k_{\Xi^0} - k) \,,
    \label{eq:fXi0}
\end{equation}
where $k_{\Xi^0}$ is the Fermi momentum of $\Xi^0$ baryon, $k_{\Xi^0} = \sqrt{ \mu_B^2 - M_{\Xi^0}^2 }$.
In Fig.~\ref{fig:phasespace_xi0}, we plot the phase-space distribution for baryons and quarks in the corresponding dual picture.
From this distribution, the number density of $\Xi^0$ and the total energy density become
\begin{align}
    \frac{\, n_{\Xi^0} \,}{g} &= \int_0^{k_{\Xi^0}} \!\!\!\!\!\! dk\, D(k)\,,
    \\
   \frac{\, \varepsilon \,}{g} &= \frac{1}{N_c^3}\int_0^{k_Y} \!\!\!\!\!\! dk\, D(k) \left[\frac{E_Y(k)}{ B_{d}^{Y} }
    - \frac{E_N(k)}{ B_{d}^{n} }\right] 
    \notag \\
    & - \left(1 - \frac{1}{ B_{d}^{n} N_c^3}\right) \int_0^{\kbu} \!\!\!\!\!\! dk\, D(k) E_N(k) 
    \notag \\
    & +  \int_0^{\ksh} \!\!\!\!\!\! dk\, D(k) E_N(k)
      + \int_0^{k_{\Xi^0}} \!\!\!\!\!\! dk\, D(k)E_{\Xi^0}(k)\,,
\end{align}
where $E_{\Xi^0}(k) = \sqrt{k^2 + M_{\Xi^0}^2}$ represents the energy for $\Xi^0$.
We find the chemical potential is
\begin{align}
    \mu_{\Xi^0} = \left(\frac{\partial \varepsilon}{\partial n_{\Xi^0}}\right)_{n_n,n_Y} = E_{\Xi^0}(k_{\Xi^0})\,.
\end{align}
The $\beta$-equilibrium condition implies $\mu_B = \mu_{\Xi^0}$, and this relation determines the value of $k_{\Xi^0}$ at a given $\mu_B$.

\begin{figure}
    \centering
\includegraphics[width=0.96\columnwidth]{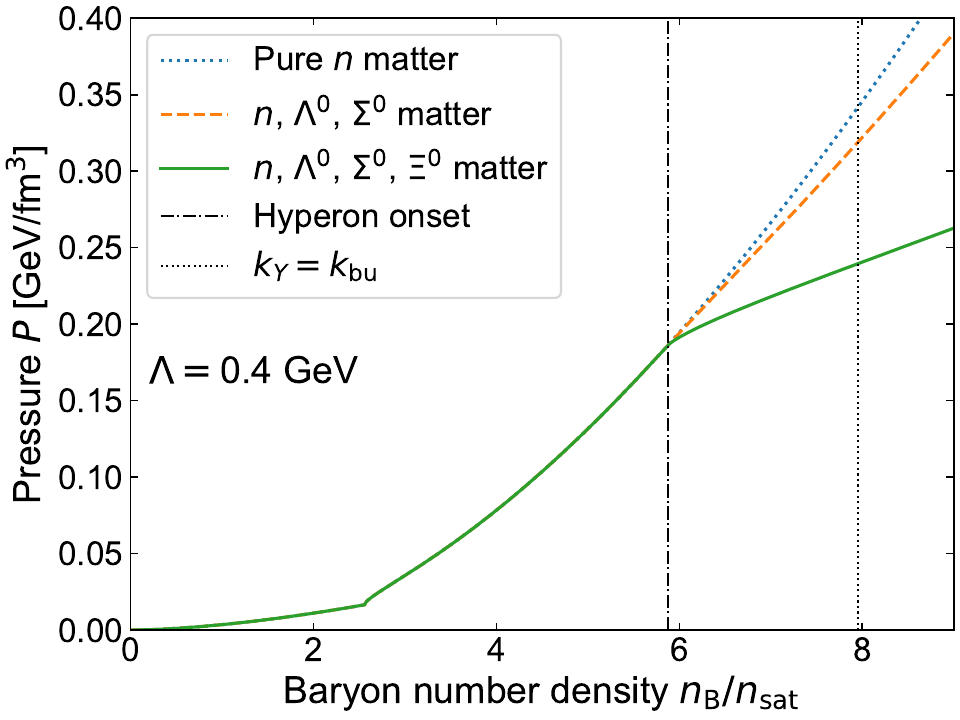}
    \caption{Comparison of the energy density at a given baryon density among the pure neutron matter, the $S=-1$ hyperon matter, and the hyperon matter with $\Xi^0$.}
    \label{fig:ne_xi0}
\end{figure}

Shown in Fig.~\ref{fig:ne_xi0} is the EoS including $\Xi^0$. We see the radical softening above the hyperon onset.
But this softening regime ends soon as $u$- or $s$-quarks get saturated within a small density interval.
The computations of $u$- and $s$-quark saturation will be presented elsewhere.

\subsection{Including heavier baryons: the case of $\Delta^0$}
\label{sec:Delta0}

Here, we discuss that $\Delta^0$, which is another charge-neutral baryon lighter than $\Xi^0$, does not appear in this setup from the vantage point of energetics.
One can compute the chemical potential for $\Delta^0$ from the relation $\mu_{\Delta^0} = (\partial \varepsilon / \partial n_{\Delta^0})_{n_n, n_Y}$:
\begin{align}
    \mu_{\Delta^0} &= E_{\Delta^0}(k_{\Delta^0}) 
    - \frac{ B_{d}^{\Delta^0} }{ B_{d}^{Y} } E_Y(k_{\Delta^0}) 
    + \frac{ B_{d}^{\Delta^0} }{ B_{d}^{Y} } \mu_Y\,,\notag \\
    &= E_{\Delta^0}(k_{\Delta^0}) - 2 E_Y(k_{\Delta^0}) + 2 \mu_Y\,,
\end{align}
where $k_{\Delta^0}$ is the Fermi momentum of $\Delta^0$ and $B_{d}^{\Delta^0} = 2/3$ is the baryon number carried by $d$ quarks inside the $\Delta^0$ baryon.
As one can explicitly verify, the value of the chemical potential is independent of $h_{\Delta^0}$.
Therefore, one can set the $\beta$-equilibrium condition regardless of the value of $h_{\Delta^0}$.
In the $\beta$-equilibrium, one finds $\mu_Y = \mu_B$ and $\mu_{\Delta^0} = \mu_B$.

The variation of the energy with respect to $h_{\Delta^0}$ reads
\begin{align}
   & \left(\frac{\partial \varepsilon}{\partial h_{\Delta^0}}\right)_{n_B} 
   \notag \\
    &= d_{\Delta^0} \frac{g}{N_c^3} \int_0^{k_{\Delta^0}} dk D(k)  \notag \\
    & ~~~~~~ \times \left[ E_{\Delta^0}(k) - \frac{ B_{d}^{\Delta^0} }{ B_{d}^{Y} } E_Y(k) - \left(1 - \frac{ B_{d}^{\Delta^0} }{ B_{d}^{Y} } \right) \mu_B \right]
    \notag \\
    &= d_{\Delta^0} \frac{g}{N_c^3} \int_0^{k_{\Delta^0}} dk D(k) \left[ E_{\Delta^0}(k) - 2 E_Y(k) + \mu_B \right] > 0\,.
\end{align}
One can easily verify that the integrand is positive for $k< k_{\Delta^0}$. From this argument, one can say that $h_{\Delta^0} = 0$ is the minimum energy solution. This means that the appearance of the $\Delta^0$ baryons is disfavored in this setup.

\subsection{Applicability of the solution}

The aforementioned solution is valid when the following four conditions are fulfilled:
(i) $\Lambda \lesssim 740~\text{MeV}$.
(ii) $k_Y < \kbu$.
(iii) $f_u(q) < 1$ and $f_s(q) < 1$ for any $q$.
(iv) $\mu_B \lesssim N_c^{3/2} \Lambda_{\rm QCD}$.

First, the ansatz for $f_n$ and $f_Y$~(\ref{eq:fn}, \ref{eq:fY}) is valid as long as the saturation of the $d$-quark distribution occurs at $\mu_B < M_Y$.
When $\ksh^Y \equiv \sqrt{\mu_B^2 - M_Y^2}$ exceeds $\kbu$, neutrons at $\ksh$ can decay into hyperons with the momentum $\ksh^Y (\ge \kbu)$.
Removing a neutron at $\ksh$ reorganizes the shell structure and reduces the energy by $\mu_B (>E_n(\ksh))$, and the baryon number conservation demands the system to accommodate another baryon.
Adding a hyperon in the shell domain costs at least the energy of $\sqrt{M_Y^2+\kbu^2}$.
When the conversion of neutrons to hyperons within the shell is energetically preferred, the $S=-1$ hyperons also develop the shell structure like in the neutron distribution. 
Such a situation may be realized when $\Lambda$ is large and the saturation of the quark distribution occurs at low $\mu_B$.
It occurs when $\Lambda \approx 740~\text{MeV}$ or larger.
This is derived from the Fermi momentum at which the saturation of the $d$-quark distribution occurs $k_{\rm sat} \approx (2/N_c)^{1/2} \Lambda$, and the condition that the $d$-quark distribution saturates below the hyperon onset $\mu_{\rm sat} = \sqrt{k_{\rm sat}^2 + M_N^2} > M_Y$.

Secondly, when $k_Y$ exceeds $\kbu$, the solution~(\ref{eq:fn_opt}, \ref{eq:fY_opt}) is no longer valid. 
One has to also consider the shell structure in $S=-1$ hyperon distribution $f_Y$ as in the neutron distribution.
This applicable limit is plotted with the dotted lines in Figs.~\ref{fig:eos04} and \ref{fig:ne_xi0}.

Thirdly, when the $u$ and $s$ quark distributions, $f_u(q)$ and $f_s(q)$, are saturated, 
the $\Xi^0$ hyperon will also have the low occupation at low momentum and the shell structure at high momentum.
Such a situation demands us to manifestly consider $u$- or $s$-quark saturation conditions.
We note that, once either $u$- or $s$-quark states are saturated together with $d$-quarks, no baryon octet can be free from the saturation conditions; all baryon octets are suppressed at low momenta.

Finally, when $\mu_B \gtrsim N_c^{3/2} \Lambda_{\rm QCD}$, the color screening suppresses the confining gluons so that Quarkyonic picture may no longer be valid.

\subsection{Toward a more realistic neutron-star equation of state}

In this paper, we have modeled the neutron star matter with only the charge-neutral baryons, and there are two important effects we have neglected here: the lepton contribution and the interaction.

Regarding the first point, a neutron star in reality is composed not only of strongly interacting particles but of leptons as well.
The charge neutrality condition together with the $\beta$-equilibrium introduces a certain fraction of protons and leptons.
Since the chemical potential of leptons is finite in such a realistic environment, the chemical potential for the charged baryon can be modified.
Therefore, the threshold for the negatively charged baryons may be shifted to lower density by the charge chemical potential $\mu_Q$.
Nevertheless, the shift in the hyperon threshold in this paper holds regardless of whether there is a finite $\mu_Q$.
As long as the $d$-quark states are saturated, there is a systematic shift in the threshold, and such shift is of the order of $\sim 0.1~\text{GeV}$ while the effect of $\mu_Q$ may be of the order of $\sim 0.01~\text{GeV}$.
So, even if $\mu_Q$ is nonzero, which can lower the charged hyperon threshold, the picture we have presented in this paper remains intact.

Regarding the second point, although we have taken into account some fraction of the interaction by taking into account the confinement effect, essentially the low-energy neutrons are treated as an ideal gas, at which the effect of the interaction is shown to be already important.
The interaction may be modeled simply by the excluded volume effect~\cite{Jeong:2019lhv, Poberezhnyuk:2023rct} as in the van-der-Waals EoS~\cite{Vovchenko:2016rkn, Fujimoto:2021dvn}.
Work in these directions is in progress.

\section{Summary and conclusion}
\label{sec:summary}

In this paper, we demonstrated the extension of ideal dual Quarkyonic (IdylliQ) model to the case of multiple flavors.
We pointed out that by considering the quark substructure of baryons, there may be an additional mechanism that may also work toward the direction of mitigating the hyperon puzzle.
This mechanism arises because, in neutron matter, one cannot insert a strangeness $S =-1$ and charge neutral $Q=0$ baryon without that baryon having a down quark in it, but in Quarkyonic Matter, the down quark Fermi sea is filled.
The first available baryon state with strangeness that is electrically neutral and has no down quark in it is the $\Xi^0$ baryon composed of two strange quarks and an up quark.
To show this mechanism clearcut, we ignored small isospin splitting in the mass, e.g., $M_{\Lambda^0} = M_{\Sigma^0}$.

The main results of this paper are two-fold:
\begin{enumerate}
\item The threshold of the $S=-1$ hyperons is shifted from $\mu_B = M_Y$ to $2 M_Y - M_N$, where $M_Y$ and $M_N$ are the mass of hyperons with strangeness $S=-1$ and nucleons, respectively.
This can be understood as follows:
In order to keep the minimum energy state with the $d$-quark distribution saturated, one has to replace one neutron, which contains two $d$'s, with two hyperons, which contains one $d$.
\item The EoS is softened mildly by $S=-1$ hyperons, but not as strongly as would be the case if the quark substructure of baryons was not properly accounted for.
The bulk of the softening in the EoS comes from the $\Xi^0$ hyperon.
\end{enumerate}

Both these points make the hyperon contribution to the EoS feeble and thus works toward the resolution of the hyperon puzzle.
Hyperons at low momenta appear in the system with a low probability and thus affect the EoS only weakly.
In this regard, a large number of hyperon species is no longer crucial because hyperons that have a $d$-quark in them suppress one another to occupy the limited phase space for $d$-quarks.
Meanwhile, hyperons at high momenta are almost free from the pre-occupied phase space, but those hyperons at high momenta not only increase energy density but also the pressure significantly, and hence do not soften the EoS much.
Once the $\Xi^0$ appears, there is a significant softening in the EoS.
Nevertheless, the shifted threshold is high enough so that it may not greatly affect the maximum mass of neutron stars.

We note that the interaction and realistic structures of baryons are missing in the aforementioned picture, therefore we do not claim that this is not yet the full solution of the hyperon puzzle.
The current mechanism can be studied in more realistic setup by integrating this into the current advanced phenomenological studies of hyperons.
We believe such directions worth further investigation, but they are left as future works.

\begin{acknowledgments}
    YF and TK would like to thank the Yukawa Institute for Theoretical Physics at Kyoto University and RIKEN iTHEMS, where part of this work was completed during the International Molecule-type Workshop ``Condensed Matter Physics of QCD 2024'' (YITP-T-23-05).
    The work of YF and LM was supported by the Institute for Nuclear Theory's U.S.\ DOE Grant No.\ DE-FG02-00ER41132.
    YF is supported by Japan Science and Technology Agency (JST) as part of Adopting Sustainable Partnerships for Innovative Research Ecosystem (ASPIRE), Grant No.\ JPMJAP2318.
    TK is supported by JSPS KAKENHI Grant No.\ 23K03377 and No.\ 18H05407 and by the Graduate Program on Physics for the Universe (GPPU) at Tohoku University.
\end{acknowledgments}

\section*{Data availability}
The data are not publicly available. The data are available from the authors upon reasonable request.

\appendix
\section{Details of calculations using Jacobian}
\label{app:Jacobi}

Here, we give details of calculations using Jacobian for derivatives that appear in the main text.

We first take up the computation of neutron chemical potential. We take the derivative of energy density holding $n_Y$ fixed. It can be written
\begin{equation}
\mu_n 
= \bigg(\frac{\partial \varepsilon }{\partial n_n } \bigg)_{n_Y}
= \frac{\partial(\varepsilon, n_Y)}{\partial(n_n, n_Y)} 
= \frac{\partial(\varepsilon, n_Y)}{\partial(\ksh, k_Y)} \bigg/
\frac{\partial(n_n, n_Y)}{\partial(\ksh, k_Y)} \,.
\end{equation}
Each determinant reads
\begin{equation}
\frac{\partial(\varepsilon, n_Y)}{\partial(\ksh, k_Y)}   
= \bigg( \frac{\partial \varepsilon }{\partial \ksh }   \bigg)_{k_Y}
\bigg( \frac{\partial n_Y }{\partial k_Y }   \bigg)_{\ksh}
-
\bigg( \frac{\partial \varepsilon }{\partial k_Y }   \bigg)_{\ksh}
\bigg( \frac{\partial n_Y }{\partial \ksh }   \bigg)_{k_Y} \,,
\end{equation}
and
\begin{equation}
\frac{\partial(n_n, n_Y)}{\partial(\ksh, k_Y)}   
= \bigg( \frac{\partial n_n }{\partial \ksh }   \bigg)_{k_Y}
\bigg( \frac{\partial n_Y }{\partial k_Y }   \bigg)_{\ksh}
-
\bigg( \frac{\partial n_n }{\partial k_Y }   \bigg)_{\ksh}
\bigg( \frac{\partial n_Y }{\partial \ksh }   \bigg)_{k_Y} \,.
\end{equation}
Noting $(\partial n_Y/\partial \ksh)_{k_Y} = 0$ since holding $k_Y$ and $h_Y$ fixed does not change $n_Y$, we find
\begin{align}
    \mu_n 
    = \left(\frac{\partial \varepsilon}{\partial \ksh}\right)_{k_Y}
    \bigg/\left(\frac{\partial n_n}{\partial \ksh}\right)_{k_Y} \,.
\end{align}
Explicitly,
\begin{align}
    \left(\frac{\partial \varepsilon}{\partial \ksh}\right)_{k_Y} 
    &=
    - D(\kbu) E_N(\kbu) \left(1 - \frac{1}{ B_{d}^{n} N_c^3 }\right) \frac{\partial \kbu}{\partial \ksh} 
    \notag \\
    &~~~ + D(\ksh) E_N(\ksh)\,,
    \notag \\
     \left(\frac{\partial n_n}{\partial \ksh}\right)_{k_Y} &=
    - D(\kbu)\left(1 - \frac{1}{ B_{d}^{n} N_c^3}\right) \frac{\partial \kbu}{\partial \ksh} + D(\ksh)\,.
\end{align}
The chemical potential for hyperons $\mu_Y$ can be computed in the same way,
\begin{align}
    \mu_Y 
    &= \frac{\partial(\varepsilon, n_n)}{\partial(\ksh, k_Y)} \bigg/ \frac{\partial(n_Y, n_n)}{\partial(\ksh, k_Y)} \notag \\
    &= \frac{\left(\frac{\partial \varepsilon}{\partial k_Y}\right)_{\ksh}}
    {\left(\frac{\partial n_Y}{\partial k_Y}\right)_{\ksh}}
    -\frac{\left(\frac{\partial n_n}{\partial k_Y}\right)_{\ksh}}
    {\left(\frac{\partial n_Y}{\partial k_Y}\right)_{\ksh}} \mu_n\,,
\end{align}
where
\begin{align}
    \left(\frac{\partial \varepsilon}{\partial k_Y}\right)_{\ksh}&= g d_Y \frac{h_Y}{N_c^3} D(k_Y) \left[ E_Y(k_Y)  
    - E_N(K_Y) \frac{ B_d^Y }{ B_{d}^{ n} }\right]\,, \\
    \left(\frac{\partial n_Y}{\partial k_Y}\right)_{\ksh} &= g d_Y  \frac{h_Y}{ N_c^3} D(k_Y)\,, \\
    \left(\frac{\partial n_n}{\partial k_Y}\right)_{\ksh} &= - g d_Y  
    \frac{B_{d}^Y}{B_{d}^{n} } \frac{h_Y}{N_c^3} D(k_Y)\,.
\end{align}
Using these we obtain the chemical potentials in the main text:
\begin{align}
    \mu_n &= \frac{- D E_N \big|_{\kbu} \left(1 - \frac{1}{ B_{d}^{n} N_c^3 }\right) \frac{\partial \kbu}{\partial \ksh}
    + D E_N \big|_{\ksh} }
    {- D(\kbu)\left(1 - \frac{1}{ B_{d}^{n} N_c^3}\right) \frac{\partial \kbu}{\partial \ksh} + D(\ksh) }\,,\\
    \mu_Y
     &= E_Y(k_Y) - \frac{B_{d}^Y}{B_{d}^{n} } E_N(k_Y) + \frac{B_{d}^Y}{B_{d}^{n} } \mu_n\,.
     \label{eq:muY}
\end{align}
Next we compute the derivative of $\varepsilon$ with respect to $h_Y$ for a fixed $n_B$:
\begin{align}
    &\left(\frac{\partial \varepsilon}{\partial h_Y}\right)_{n_B} 
    = \frac{\partial(\varepsilon, n_B)}{\partial(h_Y, n_B)} = \frac{\partial(\varepsilon, n_B)}{\partial(h_Y, \ksh)} \bigg/ \frac{\partial(h_Y, n_B)}{\partial(h_Y, \ksh)} \notag \\
    &= \frac{\left(\frac{\partial \varepsilon}{\partial h_Y}\right)_{\ksh}\left(\frac{\partial n_B}{\partial \ksh}\right)_{h_Y} - \left(\frac{\partial n_B}{\partial h_Y}\right)_{\ksh} 
    \left(\frac{\partial \varepsilon}{\partial \ksh}\right)_{h_Y}}
    { \left(\frac{\partial n_B}{ \partial \ksh}\right)_{h_Y}
    } \notag \\
    &= \left(\frac{\partial \varepsilon}{\partial h_Y}\right)_{\ksh} 
        - \left(\frac{\partial n_B}{\partial h_Y}\right)_{\ksh} 
        \bigg( \frac{\partial \varepsilon}{\partial n_B} \bigg)_{h_Y} \notag \\
    &= d_Y \frac{g}{N_c^3} \int_0^{k_Y} dk D(k) 
    \notag \\
  &~~~  \times \left[ E_Y(k) - \frac{B_{d}^{Y}}{ B_{d}^{n} } E_N(k) - \left(1 - \frac{B_{d}^{Y}}{B_{d}^{n}}\right) 
    \mu_B \right]\,.
\end{align}

\section{Expression of the equation of state}
\label{app:eos}
The equation of state (EoS) is always a function of the Fermi momentum $\ksh$.
Below the hyperon threshold, the EoS is described by the pure neutron matter with the low momentum part of the $d$ quark states saturated.
Above the hyperon threshold, the EoS is a mixture of neutrons and degenerate hyperons.

\subsubsection{Below the saturation and below the hyperon threshold}

The EoS is that of an ideal gas in this regime:
\begin{equation}
    n_B = g \frac{\ksh^3}{6\pi^2} \,,
~~~~    \varepsilon = g e_3(\ksh, M_N)\,,
\end{equation}
with $ \mu_B = \sqrt{\ksh^2 + M_N^2}$.
The function $e_3$ is defined as
\begin{align}
    e_3(k, M) & = \int_0^{k} dx\, D(x) \sqrt{k^2 + M^2}
    \notag \\
    &= \frac{1}{16\pi^2} k \sqrt{k^2 + M^2} (2 k^2 + M^2) 
    \notag \\
    &~~~+ \frac{1}{16\pi^2} M^4 \ln \left(\frac{\sqrt{k^2 + M^2} - k}{M}\right)\,.
\end{align}

\subsubsection{Saturation condition for pure neutron matter}
The saturation takes place when $f_d(q=0)=1$.
This happens when
\begin{equation}
\hspace{-0.2cm}
    f_d(q=0) = B_{d}^{n} N_c^3 \left[1 - e^{-\ksh/(N_c \Lambda)} \left(1 + \frac{\ksh}{N_c \Lambda} \right) \right]\,.
\end{equation}

\subsubsection{Above the saturation and below the hyperon threshold}

Above the saturation, one has to solve the equation to obtain the shell thickness $\Delta$ or the fermi momentum for the under-occupied bulk part $\kbu = \ksh - \Delta$.
The equation is
\begin{align}
 \hspace{-0.2cm}
\frac{N_c \Lambda + \ksh - \Delta}{N_c \Lambda + \ksh - \Delta - (N_c \Lambda + \ksh) e^{-\Delta / (N_c \Lambda)}} = B_{d}^{n} N_c^3\,.
\end{align}
Then, from $\Delta$ determined above, the EoS is obtained as
\begin{widetext}
\begin{align}
    n_B &= - g \left(1 - \frac{1}{ B_{d}^{n} N_c^3 }\right) \frac{\kbu^3}{6\pi^2}+ g \frac{\ksh^3}{6\pi^2} \,,\\
    \varepsilon &=  - g \left(1 - \frac{1}{ B_{d}^{n} N_c^3}\right) e_3(\kbu, M_N) + g e_3(\ksh, M_N)\,,\\
    \mu_B
    &= \frac{- \kbu^2 E_N(\kbu) \left(1 - \frac{1}{ B_{d}^{n} N_c^3}\right) \frac{\partial \kbu}{\partial \ksh}
    + \ksh^2 E_N(\ksh) }
    {- \kbu^2 \left(1 - \frac{1}{ B_{d}^{n} N_c^3}\right) \frac{\partial \kbu}{\partial \ksh} + \ksh^2 } \,, 
    \label{eq:app_muB}
\end{align}
\end{widetext}
where in the last line, we use the relation
\begin{align}
    \frac{\partial \Delta}{\partial \ksh} &= - \frac{N_c \Lambda \Delta}{(N_c \Lambda + \ksh)(\ksh - \Delta)}\,,\\
    \frac{\partial \kbu}{\partial \ksh} &= 1 - \frac{\partial \Delta}{\partial \ksh} = \frac{\ksh (N_c \Lambda  + \ksh - \Delta)}{(N_c \Lambda + \ksh)(\ksh - \Delta)}\,.
\end{align}

\subsubsection{Above the saturation and above the hyperon threshold}

After the $d$-quark saturation occurs and $\mu_B$ exceeds the hyperon threshold, the EoS are
\begin{widetext}
\begin{align}
    \frac{\, n_B \,}{g} &= \frac{1}{N_c^3}\left( \frac{1}{ B_{d}^{Y} } -\frac{1}{ B_{d}^{n} } \right) \frac{k_Y^3}{6\pi^2} 
   - \left(1 - \frac{1}{ B_{d}^{n} N_c^3}\right) \frac{\kbu^3}{6\pi^2} +  \frac{\ksh^3}{6\pi^2} \,,\\
    \frac{\, n_S \,}{g} &=  -\frac{1}{ B_{d}^{Y}  N_c^3} \frac{k_Y^3}{6\pi^2}\,,\\
    \frac{\, \varepsilon \,}{g} 
    &= \frac{1}{N_c^3} \left[\frac{e_3(k_Y, M_Y)}{ B_{d}^{Y} } - \frac{e_3(k_Y, M_N)}{ B_{d}^{n} }\right] 
     -  \left(1 - \frac{1}{ B_{d}^{n} N_c^3}\right) e_3(\kbu, M_N) + e_3(\ksh, M_N) \,, 
\end{align}
\end{widetext}
with $\mu_B$ same as in Eq.~\eqref{eq:app_muB}.
And the $k_Y$ is determined through the $\beta$-equilibrium condition
\begin{align}
    \mu_B = 2 E_Y(k_Y) - E_N(k_Y)\,.
\end{align}

\section{Stepwise minimization with fixed baryon density}
\label{sec:one-by-one}

\begin{figure}
    \vspace{-0.5cm}
    \centering
    \includegraphics[width=1.0\columnwidth]{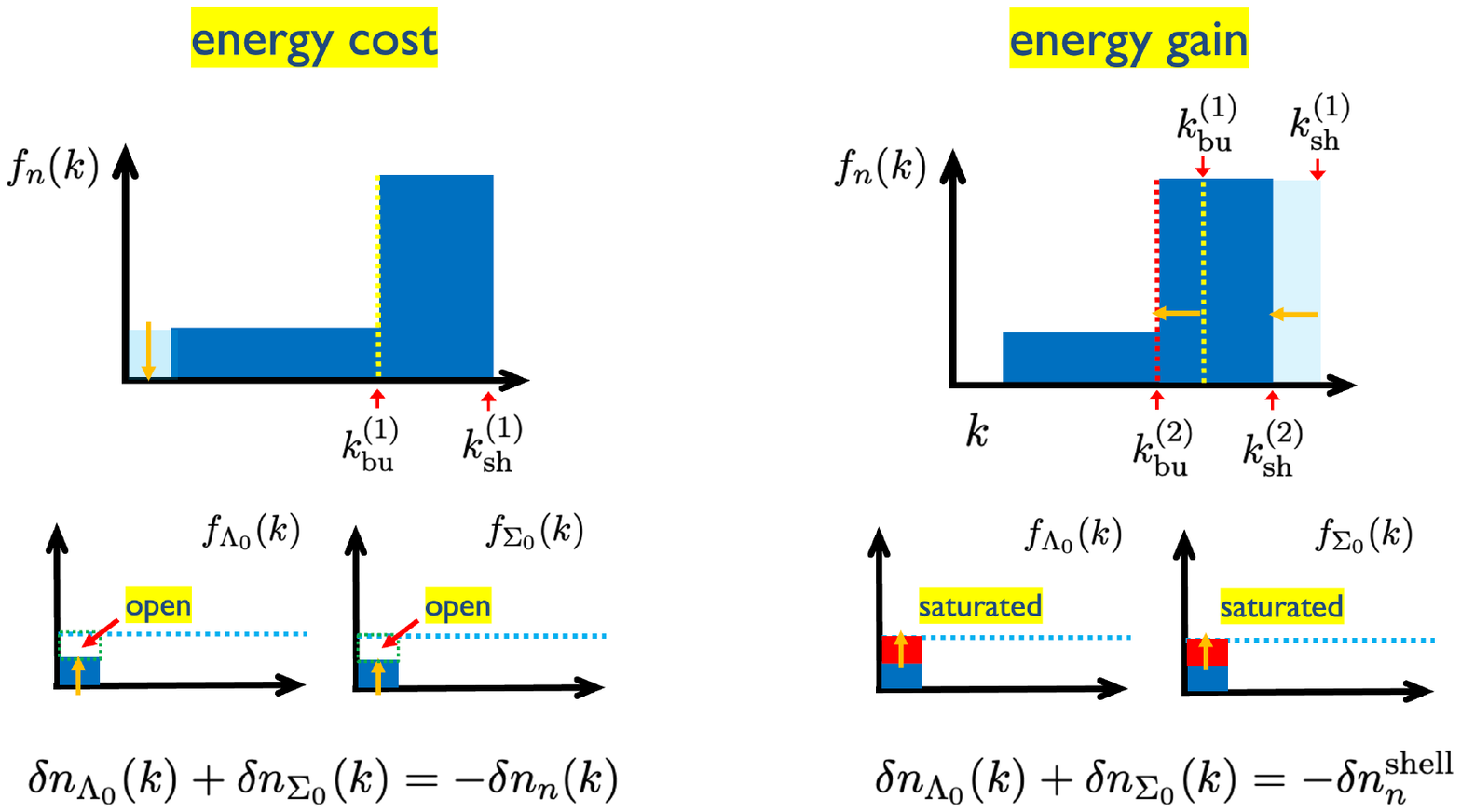}
    \vspace{-0.8cm}
        \caption{
        Compositional changes initiated with the conversion of neutrons into hyperons which in turn triggers structural changes in the neutron momentum shell. The $d$-quark states are saturated at low momenta before and after these processes. Each conversion process conserves the baryon number.
        (Left)
        Neutrons at the momenta $k$ transform into $\Lambda^0$ or $\Sigma^0$, $n(k) \rightarrow \Lambda(k)$ or $\Sigma^0(k)$ with the energetic cost. Hyperons contain fewer $d$-quarks than neutrons so that the conversion $n\rightarrow Y$ in the bulk leaves the phase space open for $d$-quarks or other hyperons; 
        (Right) The open phase space is filled by hyperons converted from neutrons in the shell. This process continues until $d$-quark states get saturated. Removing the baryon number from the shell, the shell structure is reorganized by reducing $\ksh$ and $\kbu$ as $\kbu^{(1)} \rightarrow \kbu^{(2)}$ and  $\ksh^{(1)} \rightarrow \ksh^{(2)}$. This conversion reduces the energy of the system.
        Although we show these processes sequentially, they can occur at once. Whether the energy of the system decreases or not depends on the value of $k$.}
    \label{fig:step}
\end{figure}

In this appendix we consider the stepwise variation of energy, manifestly holding $n_B$ fixed. 
We consider the conversion of $ n(k) \rightarrow Y(k)$ for baryons with momentum $k\le \kbu$ and subsequent modification of the shell.
The advantage of this strategy is that one can identify which processes cost and gain the energy,
see Fig.~\ref{fig:step}.

We first consider the process $n(k) \rightarrow \Lambda^0 (k)$ or $\Sigma^0 (k)$. The variation of each density at momentum $k$ is
\begin{equation}
\delta n_n (k) 
+ \delta n_{\Lambda^0}^{(1)} (k) + \delta n_{\Sigma^0}^{(1)} (k) 
= 0 \,,    
\end{equation}
where we defined
\begin{equation}
    \delta n_i (k) \equiv \delta f_i (k) D(k) dk \,,
\end{equation}
with $D(k)=k^2/2\pi^2$ being the density of states at $k$.
Assuming $\Lambda^0$ and $\Sigma^0$ are generated at equal probability, in this first conversion process the density of each hyperon species changes as
\begin{equation}
    \delta n_Y^{(1)}(k) \equiv \delta n_{\Lambda^0}^{(1)} (k) = \delta n_{\Sigma^0}^{(1)} (k)
    = - \frac{\delta n_n(k)}{2} \,.
\end{equation}
After this conversion, the $d$-quarks are no longer saturated. 
This is because we lose a single $d$-quark for each $n(k;udd) \rightarrow Y(k;uds)$ process. We also note that this first process costs the energy $E_Y(k) - E_n(k)$ for each conversion and never occurs unless another energy reduction process becomes possible by this conversion.

To fill the hole in the $d$-quark states in the bulk, we consider the second conversion process in which the neutrons in the shell decay into hyperons in the bulk. In this second process, the hyperon densities increase until the $d$-quark states are saturated,
\begin{equation}
    \delta n_Y^{(2)}(k) = \delta n_{\Lambda^0}^{(2)} (k) = \delta n_{\Sigma^0}^{(2)} (k)
    = \delta n_Y^{(1)} (k) \,.
\end{equation}
Thus, the amount of decaying neutrons is
\begin{equation}
    - \delta n_n^{\rm shell}
    =  \delta n_{\Lambda^0}^{(2)} (k) + \delta n_{\Sigma^0}^{(2)} (k) 
    = - \delta n_n (k) \,. 
\end{equation}
Reducing the number of neutrons in the shell, the shell structure is reorganized until the optimized (saturated) distribution is achieved again.
As we have explicitly calculated in isospin symmetric matter, the momenta $\kbu$ and $\ksh$ for the saturated momentum shell are uniquely related.  
The $\kbu$ and $\ksh$ before (index 1) and after (index 2) satisfy the relation
\begin{equation}
    F_{\rm sat} (\kbu^{(1)}, \ksh^{(1)}) 
    = F_{\rm sat} (\kbu^{(2)}, \ksh^{(2)}) = 0 \,.
\end{equation}
and hence the relation between $d \kbu = \kbu^{(2)}-\kbu^{(1)} (<0)$ and $d \ksh = \ksh^{(2)}-\ksh^{(1)} (<0)$ is also uniquely specified.
Now we can write the variation of neutron number in the shell as
\begin{equation}
\delta n_n^{\rm shell}
= -\bigg( 1 - \frac{1}{\, B_{d}^{n} N_c^3 \,} \bigg) D(\kbu) d \kbu
+ D(\ksh) d \ksh \,.
\end{equation}
Here we assumed that the bulk just below $\kbu$ is saturated by neutrons; in the interval $[\kbu-d\kbu, \kbu]$ the amplitude increases from $1/B_{d}^{n} N_c^3$ to $1$, while the amplitude in $[\ksh-d\ksh, \ksh]$ is reduced from $1$ to $0$.

Now we examine how the energy density changes. Since we write the variations within infinitesimal phase space, it is sufficient to multiply the energy at each phase space,
\begin{align}
\delta \varepsilon^{(1+2)} 
&= \sum_{Y=\Lambda^0, \Sigma^0} E_Y(k) \delta n_Y^{(1+2)} (k) 
\notag \\
&~~~~~ + E_n (k) \delta n_n(k)
    +  \delta \varepsilon_n^{\rm shell}
\notag\\
&= \delta n_n (k) \bigg( 
 - 2 E_Y(k) + E_n (k) 
    + \frac{ \delta \varepsilon_n^{\rm shell} }{ \delta n_n^{\rm shell} }
    \bigg)
\end{align}
where in the last step we used 
$\delta n_Y^{(1+2)} (k) = - \delta n_n(k)$ and $\delta n_n^{\rm shell} = \delta n_n(k)$. Here $\delta \varepsilon_n^{\rm shell}$ is
\begin{equation}
\delta \varepsilon_n^{\rm shell}
= - E_n D \big|_{\kbu} \bigg( 1 - \frac{1}{\, B_{d}^{n} N_c^3 \,} \bigg)  d \kbu
+ E_n D \big|_{\ksh} d \ksh \,.
\end{equation}
At chemical equilibrium, the expression $\delta \varepsilon_B^{\rm shell}/\delta n_B^{\rm shell}$ corresponds to the neutron chemical potential as shown in the main text.
We recall that $\delta n_n (k) < 0$ corresponds to the increase of $f_Y(k)$. Looking at the coefficient of $\delta n_n (k)$, we conclude that increasing $f_Y(k)$ reduces the energy when
\begin{equation}
2 E_Y(k) 
-
E_N (k)
\le \frac{\, \delta \varepsilon_B^{\rm shell} \,}
    {\, \delta n_B^{\rm shell} \,}
    \,.
\end{equation}
In the main text, we have seen that the LHS is an increasing function. Hence the inequality holds (i.e., hyperons appear) only at low momenta.
Hence we conclude that the distributions $f_n$ and $f_Y$ take the form
\begin{align}
    \label{eq:fn_ap}
    f_n(k) &= \frac{3}{\, 2 N_c^3 \,} \Theta(k - k_Y) \Theta(\kbu - k)  
    \notag \\
    &~~~ + \Theta(k - \kbu) \Theta(\ksh - k)\,,\\
    f_Y(k) &= \frac{3}{\, 2 N_c^3 \,} \Theta(k_Y - k)\,,
    \label{eq:fY_ap}
\end{align}
where $k_Y$ is defined to be
\begin{equation}
2 E_Y(k_Y) 
=
E_N (k_Y)
+ \frac{\, \delta \varepsilon_B^{\rm shell} \,}
    {\, \delta n_B^{\rm shell} \,}
    \,.
 \label{eq:min_for_e}   
\end{equation}
Since the distribution has the lowest energy, there should be no further conversion. Hence this condition to determine $k_Y$ is equivalent to the $\beta$-equilibrium condition. One can manifestly compute $\mu_n$ and $\mu_Y$ using Jacobian method and check that $\mu_n = \mu_Y$ indeed holds. The results coincide with those in the main text.

\bibliography{quarkyonic}

\end{document}